\documentclass[pdflatex,sn-mathphys-num]{sn-jnl}% 

\usepackage{graphicx}%
\usepackage{multirow}%
\usepackage{amsmath,amssymb,amsfonts}%
\usepackage[ruled,linesnumbered,noend]{algorithm2e}
\usepackage{pifont}

\usepackage{mathrsfs}%
\usepackage[title]{appendix}%
\usepackage{xcolor}%
\usepackage[table]{xcolor}
\usepackage{textcomp}%
\usepackage{graphicx}
\usepackage{subcaption}
\usepackage{hyperref}
\usepackage{caption}
\usepackage{listings}
\usepackage{tabularx}

\usepackage{manyfoot}%
\usepackage{booktabs}  % for \toprule etc.
\usepackage{array}     % for p{} column types

\theoremstyle{thmstyleone}%
\theoremstyle{thmstyletwo}%

\theoremstyle{thmstylethree}%

\begin{document}

\title[BRIDG-ICS: AI-Grounded Knowledge Graphs for Intelligent Threat Analytics in Industry~5.0 Cyber--Physical Systems]{BRIDG-ICS: AI-Grounded Knowledge Graphs for Intelligent Threat Analytics in Industry~5.0 Cyber--Physical Systems}

\author*[1]{\fnm{Padmeswari} \sur{Nandiya}}
\email{p.nandiya@ecu.edu.au}

\author[1]{\fnm{Ahmad} \sur{Mohsin}}
\email{a.mohsin@ecu.edu.au}

\author[1]{\fnm{Ahmed} \sur{Ibrahim}}
\email{ahmed.ibrahim@ecu.edu.au}

\author[1]{\fnm{Iqbal H.} \sur{Sarker}}
\email{m.sarker@ecu.edu.au}

\author[1]{\fnm{Helge} \sur{Janicke}}
\email{h.janicke@ecu.edu.au}

\affil[1]{\orgdiv{Centre for Securing Digital Futures}, \orgname{Edith Cowan University}, \orgaddress{\city{Perth}, \postcode{6027}, \state{WA}, \country{Australia}}}

% \affil[1]{\orgdiv{Centre for Securing Digital Futures}, \orgname{Edith Cowan University}, \orgaddress{\city{Perth}, \postcode{6027}, \state{WA}, \country{Australia}}}

\abstract{
Industry 5.0’s increasing integration of IT and OT systems is transforming industrial operations but also expanding the cyber–physical attack surface. Industrial Control Systems (ICS) face escalating security challenges as traditional siloed defenses fail to provide coherent, cross-domain threat insights. We present BRIDG-ICS (BRIDge for Industrial Control Systems), an AI-driven Knowledge Graph (KG) framework for context-aware threat analysis and quantitative assessment of cyber resilience in smart manufacturing environments. BRIDG-ICS fuses heterogeneous industrial and cybersecurity data into an integrated Industrial Security Knowledge Graph linking assets, vulnerabilities, and adversarial behaviors with probabilistic risk metrics (e.g. exploit likelihood, attack cost). This unified graph representation enables multi-stage attack path simulation using graph-analytic techniques. To enrich the graph’s semantic depth, the framework leverages Large Language Models (LLMs): domain-specific LLMs extract cybersecurity entities, predict relationships, and translate natural-language threat descriptions into structured graph triples, thereby populating the knowledge graph with missing associations and latent risk indicators. This unified AI-enriched KG supports multi-hop, causality-aware threat reasoning, improving visibility into complex attack chains and guiding data-driven mitigation. In simulated industrial scenarios, BRIDG-ICS scales well, reduces potential attack exposure, and can enhance cyber–physical system resilience in Industry 5.0 settings.}

\keywords{Smart Manufacturing, Threat Modelling, Knowledge Graph, Risk Assesment, Industrial Control System (ICS) }



\maketitle

\section{Introduction}\label{sec1}

Industrial Control Systems (ICS) in Operational Technology (OT) environments manage essential processes across 
sectors such as energy, water, and transportation~\cite{hassan2024systematic}. As these 
systems evolve within the broader vision of Industry~5.0, characterized by tighter 
human--machine collaboration and increasing use of artificial intelligence, their 
Connectivity requirements have grown. ICS components that were once isolated now routinely 
interface with enterprise Information Technology (IT) systems and cloud-based platforms to enable data sharing, 
remote access, and advanced analytics~\cite{sun2025unlocking}. This gradual integration, 
Accelerated by the proliferation of Industrial IoT devices, enhances operational efficiency 
but simultaneously expands the attack surface and weakens the traditional OT ``air-gap'' 
security model~\cite{hassan2024systematic,tyagi2021itotconvergence,srinivasan2025threatbasedsecuritycontrolsprotect}, 
exposing industrial environments to new categories of cyber threats.
 As a result, industrial cybersecurity continues 
to lag behind mainstream IT security practices, with legacy systems relying on 
outdated protection mechanisms despite increasingly frequent and sophisticated attacks~
\cite{anton2019devildetailattackscenarios}. These developments highlight a growing gap 
between modern operational demands and the security posture of existing OT infrastructures. 
As Industry 5.0 advances towards deeper cyber-physical integration and more seamless 
human–machine collaboration, cybersecurity is fundamental to ensuring 
resilient and sustainable industrial operations, helping to manage the amplified risks 
introduced by IT/OT convergence and the expanding Industrial IoT ecosystem~
\cite{frontiers2022sustainability,sun2025unlocking}.

The heightened exposure introduced by Industry 5.0 connectivity is reflected in the current industrial cybersecurity trends. According to the Kaspersky ICS-CERT Q2 2025 report, nearly one-third of the monitored OT systems experienced malicious activity, indicating a continued rise in attacks on industrial infrastructure, particularly within the manufacturing sector~\cite{kaspersky2025industrialincidents}. High-impact incidents further illustrate the severity of these threats. The Stuxnet operation showed that cyber intrusions can cause physical damage to industrial equipment, and the \textit{dr0wned} study demonstrated that tampering with additive manufacturing processes can lead to critical component failures~\cite{belikovetsky2017dr0wned}. In addition, more than sixty ransomware events in recent years have halted manufacturing operations by exploiting vulnerabilities introduced through IT and OT integration~\cite{ALQUDHAIBI2025100067}. Taken together, these developments highlight how digital attacks increasingly produce tangible physical consequences in connected Industry 5.0 environments.

\begin{figure*}[b]
  \centering
  \includegraphics[width=0.75\textwidth]{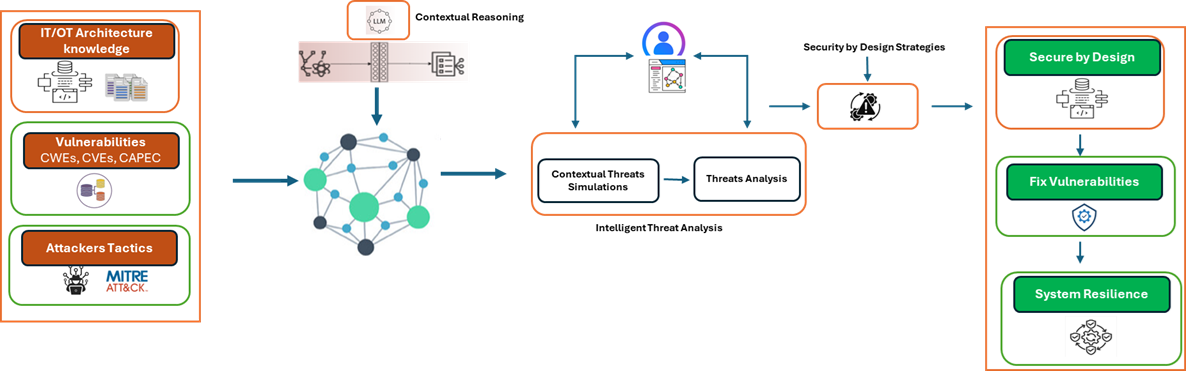}
  \caption{Intelligent threat analysis and Security-by-Design workflow integrating Knowledge Graph reasoning, LLM-based contextual analysis and system resilience assessment.}
  \label{fig:intelligent-threat-analysis}
\end{figure*}

As these incidents illustrate, understanding how adversaries infiltrate and move through industrial networks is essential. Attackers frequently exploit the interface between IT and OT systems, using vulnerable IT/IoT assets to pivot to controllers and safety-critical devices within the operational environment~\cite{MILLER2021100464, Anton_2021}. This pathway enables        
multi-stage intrusions that can bypass traditional perimeter-focused defenses, yet many organizations still lack the situational awareness needed to track how such attacks unfold across interconnected domains~\cite{electronics13050917}.

A further obstacle to effective defense is the fragmented nature of industrial threat intelligence. Indicator feeds typically provide isolated alerts without linking adversarial behavior to underlying system dependencies~\cite{nankya2023securing}. Structured ontologies such as MITRE ATT\&CK partially address this issue by offering a unified vocabulary for adversarial tactics and techniques across IT and OT~\cite{10.1145/3687300,9638634, mitre_ics}. When integrated into knowledge-graph representations that capture assets, vulnerabilities, and operational processes, these ontologies enable the reconstruction of cross-domain attack paths and clarify how cyber events can escalate into physical consequences~\cite{11143218}. This combined view supports coherent, context-aware modelling of multi-stage threats in Industry 5.0 environments.

While knowledge-graph approaches offer greater structure than traditional threat-intelligence feeds, they still fail to model how cyber–physical attacks unfold in industrial environments. Existing methodologies such as STRIDE, PASTA, ATT\&CK matrices, risk-scoring models, and anomaly-detection techniques address specific dimensions of the threat landscape, yet none provides an integrated framework that links the industrial context, vulnerability information, and adversarial behavior into a unified analytical model~\cite{technologies11060161, app14188398}. Although standardized taxonomies improve semantic consistency, most representations still offer limited support for reasoning across both cyber and operational layers~\cite{ALANEN2022108270, KHALIL2024103543}. As a result, they often overlook the process dependencies and system interactions that define ICS environments, constraining the ability to reconstruct multi-stage attack paths or evaluate the broader impact of the security controls.

Although these limitations are most visible in modelling approaches, similar challenges arise in industrial risk-assessment practices. Many techniques struggle with scalability and cannot accommodate the complexity of modern ICS environments~\cite{hassan2024systematic, KHALIL2024103543}. Conventional threat-intelligence tools further compound the problem by overlooking relationships between industrial indicators that are essential for reconstructing multi-stage attacks~\cite{cyberinitiative2021icskg}. These gaps prevent analysts from forming a coherent picture of how threats emerge and propagate across interconnected IT and OT systems.

These limitations underscore the need for a unified framework that bridges IT–OT semantics, embeds domain knowledge, and supports context-aware reasoning across attack paths to strengthen situational awareness in Industry~5.0 environments. This paper introduces \textit{BRIDG-ICS}, an AI-driven system for advanced threat modelling and improved cyber-resilience in Industry~5.0 cyber–physical systems. The approach integrates heterogeneous cybersecurity sources, CVE, CWE, CAPEC (vulnerability taxonomies), and MITRE ATT\&CK (adversarial tactics and techniques), with Industry~5.0 architectural knowledge into a unified Knowledge Graph (KG). This KG is then enriched using pre-trained Large Language Models (LLMs) and graph-embedding techniques. LLMs are employed to extract missing entities, normalize heterogeneous terminology, and infer latent semantic relations across vulnerability descriptions, industrial device functions, configuration states, and attacker techniques. Complementarily, graph embeddings reinforce and generalize these inferred relations through similarity-based link prediction, strengthening the structural coherence of the KG. This dual enrichment process reveals multi-layer dependencies, such as cross-domain exploit chains, tactic–technique mappings, and subsystem-specific attack surfaces that text-centric or manually curated tools miss. The resulting KG provides an interconnected, semantically consistent view of industrial assets, communication paths, and configuration states, with each element mapped to relevant vulnerabilities, adversary behaviors, and defensive controls. BRIDG-ICS enables automated reasoning over multi-stage cyber–physical attacks, including propagation path inference, likelihood-weighted exploit progression, and comparison of mitigation strategies, unifying IT/OT knowledge, behavioral patterns, and adversarial intent into a single reasoning pipeline. As shown in Figure~\ref{fig:intelligent-threat-analysis}, the workflow integrates contextual threat simulation, multi-hop analytical reasoning, and secure-by-design assessment, ensuring alignment between evolving adversarial behaviors and proactive defensive strategies. This establishes a scalable foundation for incorporating real-time industrial telemetry, continuous threat intelligence, and future AI-driven decision-support, enabling autonomous, context-aware, and self-defending Industry~5.0 ecosystems. The contributions of this paper are as follows:
\vspace{-2pt}
\begin{enumerate}
   \item \textit{Industry~5.0 Knowledge Graphs:} BRIDG-ICS constructs domain-specific KG by fusing industrial data with a variety of cybersecurity datasets\footnote{These datasets (CVE, CWE, CAPEC, MITRE ATT\&CK) are integrated and processed to develop knowledge graphs that are linked with local industrial contorl systems datasets. }, thereby encoding process-level semantics and dependencies among devices to support advanced threat analysis and other cybersecurity, related activities.
\vspace{-2pt}
\item \textit{Intelligent KG Enrichment:}
The approach employs LLMs for \textit{NER, RE, and NL2KG} to semantically enrich KG nodes and relations, while graph-embedding and graph-analytics methods (e.g., similarity search, clustering, and path-based metrics) infer and strengthen latent connections, uncovering deeper structural and semantic patterns across threat intelligence.

\vspace{-2pt}
\item \textit{Context-aware Threat Analysis:} The approach unifies IT and OT security views for multi-layer threat reasoning, improving insight into adversarial intent, impact propagation, and cross-domain attack chains.
\vspace{-2pt}

\item \textit{Data-Driven Attack Path Discovery:} BRIDG-ICS uses graph-based learning to simulate attack scenarios, assigning probabilistic risk attributes to predict likely attack paths and assess control effectiveness.
\vspace{-2pt}
\item \textit{Resilience and Adaptive Defense:} The proposed approach enables scenario-based simulations for adaptive defense, continuously integrating new analytics and threat intelligence to support autonomous, sustainable, collaborative Industry 5.0 ecosystems.
\vspace{-10pt}
\end{enumerate}.
\vspace{-7pt}

\section{Background}

This section reviews the fundamental elements relevant to our work: the complexity of cyber–physical ICS (Sect.~2.1), the opportunities and limitations of AI-driven threat intelligence (Sect.~2.2), and the use of knowledge graphs for unified modelling across cyber and physical layers (Sect.~2.3), forming the basis for the BRIDG-ICS framework.

\subsection{Cyber--Physical Systems Complexity and Threat Modelling}

In Industry~5.0, ICS integrate diverse components such as PLCs, sensors, and SCADA servers~\cite{nist80082,isa62443,nankya2023securing}, creating significant variability in firmware and security configurations across vendors. Unlike IT systems, ICS are tightly coupled to physical processes and must run continuously, limiting patching. As a result, vulnerabilities often persist across operational layers and supply chains~\cite{ics-myth,makrakis2021}. Assessing these weaknesses is difficult due to architectural complexity, and traditional risk models often fail to capture how vulnerabilities propagate within interconnected control environments~\cite{song2024security}. AI-enabled automation and digital twins further increase these dependencies~\cite{frontiers2024cybersecurity}. Conventional IT-centric methods (e.g., STRIDE, DREAD) do not adequately represent cyber--physical interactions \cite{Thodelling2024}, underscoring the need for frameworks that integrate cyber and physical semantics to ensure reliability and safety~\cite{KHALIL2024103543}.

\begin{figure*}[b]
\centering \includegraphics[width=0.85\textwidth]{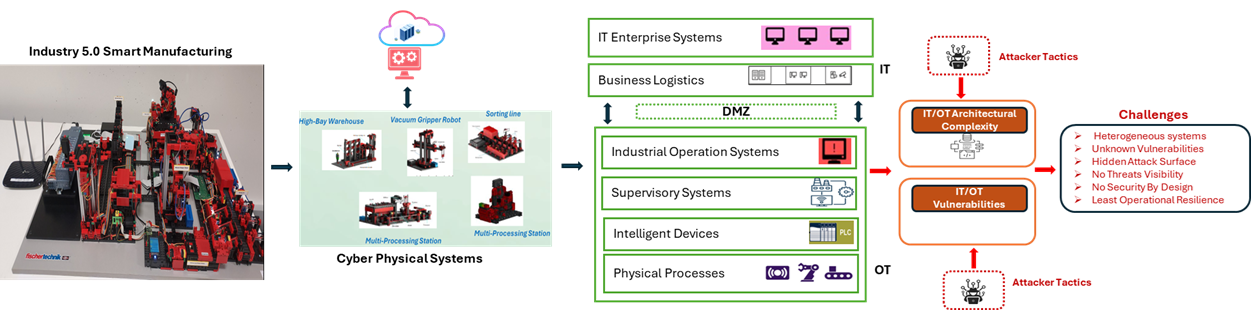} 
\caption{Industry 5.0 Smart Manufacturing integrates CPS and IT/OT layers, highlighting complexity and cybersecurity challenges.} 
\label{fig:CPS-IT-OT-complexity} 
\end{figure*} 

Figure~\ref{fig:CPS-IT-OT-complexity} shows a smart manufacturing CPS where IT and OT layers converge through multi-level connectivity. This integration, from enterprise systems to field controllers, creates dense data flows and interdependencies that expand the attack surface and expose previously hidden vulnerabilities. These challenges underscore the need for knowledge-driven, model-based approaches such as \textit{BRIDG-ICS}, which capture relationships among assets, vulnerabilities, and operational processes to enable proactive, context-aware threat analysis and resilience engineering.

\subsection{AI-Driven Analytics and Threat Intelligence}

Artificial Intelligence (AI) and analytics are increasingly integrated into OT, industrial IoT, and manufacturing execution systems. These technologies generate large volumes of threat intelligence, including CVE/CWE/CAPEC reports, ATT\&CK mappings, anomaly logs, and supply-chain telemetry; yet much of this data remains fragmented and weakly contextualized~\cite{DBLP:journals/corr/abs-2103-03530}. Although AI improves anomaly detection and predictive analysis, it cannot explain how cyber intrusions evolve into physical disruptions without awareness of process semantics and asset dependencies. Achieving genuine cyber resilience, therefore, requires analytical frameworks that combine AI-driven inference with industrial context, transforming raw indicators into actionable and interpretable threat intelligence. This requirement underpins the design of our proposed approach.

\subsection{Knowledge Graphs for Context-Aware Threat Modelling}

Effective cyber-physical threat reasoning requires representations capable of capturing both cyber and operational semantics. KG offers this capability by structuring entities and relationships into machine-interpretable graphs that support inference across heterogeneous domains~\cite{hoogan,ji2022survey}. In cybersecurity, KGs unify system assets, vulnerabilities, attack techniques, and mitigations into a shared semantic model~\cite{li2023,gao_threatkg,attackg}, formally represented as:
\[
\mathcal{G} = (\mathcal{V}, \mathcal{E}, \phi, \psi),
\tag{1}
\]
where $\mathcal{V}$ denotes entities, $\mathcal{E}$ relations, and $\phi,\psi$ their semantic mappings.

A structured threat ontology is central to modeling adversarial behavior. miter ATT\&CK provides such a foundation by cataloging tactics and techniques from real intrusions, enabling system components, vulnerabilities, and attacker actions to be represented within a unified semantic space~\cite{11143218}. When integrated into a KG, ATT\&CK supports multi-stage attack-path identification across IT and OT layers and clarifies how adversaries progress from initial compromise to physical impact.

KG-based approaches in industrial cybersecurity still rely on static taxonomies or correlation-based indicator mappings~\cite{aptkg,attackg_plus} and thus cannot model dynamic causal relations among assets, vulnerabilities, process behaviors, and adversarial objectives. \textit{BRIDG-ICS} overcomes this by integrating IT/OT asset data, operational semantics, and threat intelligence into a unified KG that supports probabilistic, multi-stage attack-chain reasoning. By embedding contextual intelligence, linking vulnerabilities, control logic, data flows, and ATT\&CK-derived adversarial behavior, the framework offers explainable traceability from digital compromise to physical consequence, enabling proactive resilience, informed mitigation, and improved situational awareness for secure Industry~5.0 systems~\cite{sikos2023cybersecurity,ics_sec,KHALIL2024103543}.

\subsection{Key Definitions}

In this work, the terms \textit{Product}, \textit{Asset}, and \textit{I5\_Asset} are used interchangeably to denote vendor-specific industrial components. According to the United States Cybersecurity and Infrastructure Security Agency (CISA), each vendor supplies its own industrial product, such as Siemens S7 PLCs, Schneider Modicon controllers, or Rockwell ControlLogix devices, representing distinct components within an industrial ecosystem~\cite{icsa}. Throughout this work, these vendor-specific components are collectively referred to as \textit{Products}.

Under the MITRE ATT\&CK for ICS framework, an \textit{Asset} refers to a functional category of an industrial component. Examples include Workstations, HMIs, Control Servers, PLCs, RTUs, IEDs, Data Historians, Application Servers, Data Gateways, Safety Controllers, VPN Servers, Jump Hosts, Field I/O devices, and Routers~\cite{mitre_ics}. These classifications emphasize functional roles rather than vendor identity.

Within BRIDG-ICS, the \textit{I5\_Asset} maps each vendor-specific \textit{Product} to its corresponding ATT\&CK-aligned \textit{Asset} category in the KG, ensuring consistent semantic representation across diverse industrial environments.

\section{Related Work}
\label{sec:related_work}

This section situates our work within existing research on cybersecurity knowledge 
representation and threat reasoning. We review prior efforts in applying KG to cybersecurity, examine ontology-driven approaches for structuring cyber–physical 
security knowledge, and analyze AI-based threat modeling techniques.

\subsection{Knowledge Graphs in Cybersecurity}

Knowledge Graphs (KGs) are widely used in cybersecurity for knowledge representation and reasoning, helping to address challenges from complex network environments, diverse attack tactics, and heterogeneous multi-source data. Increasingly sophisticated cyber threats demand cognitive intelligence technologies for deeper analysis and situational awareness. Cybersecurity information comes from numerous open and hidden sources, including vulnerability databases, Cyber Threat Intelligence (CTI) reports, social media, blogs, and the dark web~\cite{ZHAO2024103524}.

Prior studies for building cybersecurity KGs converge into two methodological streams. The first stream employs data-driven techniques (e.g., Named Entity Recognition and Relation Extraction) to automatically construct KGs from unstructured text sources like threat reports and advisories~\cite{ctikg2022, vulkg2023, cvekg2023, SARHAN2021107524, MOUICHE2025104120}. The second stream uses ontology-based modeling derived from structured datasets and standardized vocabularies to build a KG with well-defined semantics~\cite{JIA201853, ics_sec, buildkg}. Each approach offers complementary advantages: data-driven NER/RE methods can extract richer, emerging entities and relationships from text but often yield less consistent or noisier results, while ontology-driven methods provide higher semantic precision and interoperability but may lack contextual depth or miss novel threat information not covered by existing schemas. Advancing upon these foundations, recent work examines how KG support multiple stages of the threat intelligence lifecycle, from threat discovery and attack investigation to assessment, attribution, and incident correlation~\cite{ctikg2022, attackg, attackg_plus, vulkg2023, aekg4apt, electronics11152287}. These efforts underscore the growing value of KGs for automated reasoning and inference across heterogeneous cybersecurity data sources.

\renewcommand{\arraystretch}{1.15}
\setlength{\tabcolsep}{3.5pt}

\begin{table*}[h!]
\centering
\footnotesize
\caption{Comparative analysis of cybersecurity Knowledge Graph (KG) studies with our proposed BRIDG-ICS approach}
\label{tab:kg_comparison}

\begin{tabular}{@{}p{2cm} p{2cm} p{2cm} p{2cm} p{0.8cm} p{3cm}@{}}
\toprule
\textbf{Work (Year)} &
\textbf{Data Source} &
\textbf{Methodology} &
\textbf{Application} &
\textbf{Dom.} &
\textbf{Reasoning} \\
\midrule

CTI-KG (2022) & CTI feeds & NER / RE & Threat discovery & IT & Entity Linking Reasoning \\
AttacKG+ (2023) & Incident reports & Ontology + rules & Attack investigation & IT & Rule-Based Threat Reasoning \\
%VulKG (2023) & NVD + ExploitDB & Ontology fusion & Vulnerability analysis & IT & Static Vulnerability Reasoning \\
AEKG4APT (2024) & APT reports & Temporal KG embedding & APT campaign study & IT & Temporal Threat Sequence Reasoning \\  
MER-GCN (2025) & APT datasets & GCN over ICS KG & Attacker modelling & OT & Graph-Learning Threat Reasoning \\
Explainable KG (2025) & Control logs & Embedding + XAI & Process anomaly analysis & OT & Explainable Anomaly Reasoning \\  

\midrule
\rowcolor{gray!13}
\textbf{BRIDG-ICS (2025)} &
I5.0, MITRE ICS, CVE, CWE, CAPEC \& CTI &
Ontology + LLM enrichment + GDS analytics &
Proactive Threat Simulation \& Analysis &
\textbf{IT--OT} &
\textbf{NER--RE--NL2KG Semantic Threats Reasoning} \\
\bottomrule
\end{tabular}

\end{table*}

As shown in Table~\ref{tab:kg_comparison}, the majority of IT-focused KG research targets cyber knowledge graphs for enterprise environments (e.g., vulnerabilities, exploits, tactics), while more recent ICS/OT-oriented work additionally embeds physical process information and control system context. However, none of the current approaches provide a unified representation that comprehensively integrates both dimensions. Attack-path analysis is crucial for understanding how adversaries traverse interconnected assets in cyber–physical environments~\cite{fujimoto2024}, yet applying KGs to this problem across both IT and OT remains underexplored. IT-oriented graph models typically emphasize network topology or vulnerability chaining but overlook the semantic relationships and process interdependencies that drive multi-stage industrial attacks. Conversely, ICS-specific KGs focus on static ontology construction and asset inventory integration~\cite{shen2020, ics_sec, mergcn}, offering limited support for reasoning about dynamic attack progression. Developing KG-based frameworks that dynamically model, reason about, and predict adversarial behavior across IT and OT therefore remains an open challenge, directly motivating our approach.

\subsection{Ontologies in Cybersecurity}

Ontologies form the semantic backbone of knowledge graphs by defining the formal structure of entities, relationships, and constraints that enable consistent reasoning and data interoperability. In the cybersecurity domain, several ontologies have been proposed to standardize core concepts. The Unified Cybersecurity Ontology (UCO)~\cite{uco2018}, the Structured Threat Information eXpression (STIX) ontology~\cite{stix2017}, and the Security Ontology (SO)~\cite{soldatos2019} formally represent vulnerabilities, assets, attacks, and mitigations in a machine-interpretable format. They provide common vocabularies for cyber threats and defenses, enabling integration of data from sources such as CVE, CWE, attack patterns, and incident reports.

In industrial contexts, ontological modelling has been expanded to capture Operational Technology (OT) and control-system semantics that fall outside the scope of traditional IT security ontologies. The Industrial Ontology Foundry (IOF) provides a foundational schema for representing manufacturing and operational processes in smart-industry settings~\cite{iof2020}. Security ontologies derived from the ISA/IEC 62443 standard~\cite{s25030728} also formalize key industrial cybersecurity concepts—such as zones, conduits, and safety levels—that are essential for describing ICS network segmentation and trust boundaries. Additional domain-specific structure is offered by the ICS-SEC KG framework~\cite{ics_sec}, which links industrial assets and processes with cybersecurity events to support reasoning about ICS-related threats. However, ICS-SEC primarily targets IT–OT data integration and semantic validation (via SHACL) and does not support dynamic attack simulation or provide extensive evaluation in live or high-fidelity environments. Overall, existing IT and ICS ontologies provide important semantic foundations, but they remain insufficient for modeling the time-varying, context-dependent behavior of advanced attacks in Industry 4.0/5.0 systems.

\subsection{AI-Driven Threat Analysis and Comparison to Our Work}

Researchers are addressing the limitations of static threat modeling approaches through AI-driven methods, with a growing emphasis on integrating knowledge graphs and machine learning to analyze complex attack patterns~\cite{sikos2023cybersecurity, ctikg2022}. Representing cyber data as graphs enables neural models to capture hidden relationships, behavioral similarities, and multi-step adversarial strategies. For example, HINTI model employs a heterogeneous graph convolutional network to enhance threat report analysis by identifying contextual dependencies among indicators~\cite{259697}. In a related direction, CSKG4APT framework constructs a structured security knowledge graph and applies graph neural networks to infer multi-stage APT behaviors~\cite{9834133}. By combining knowledge graph embeddings with GCN-based classification, CSKG4APT predicts likely attack transitions and narrows the reasoning space for campaign reconstruction, leading to improved accuracy and interpretability.

The increasing sophistication of these AI-based models has also highlighted the importance of explainable mechanisms that reveal which threat indicators, graph structures, or semantic relationships influence model outputs \cite{yan2022explainable, mendes2023explainableartificialintelligencecybersecurity}. Such transparency is essential for analyst confidence and operational validation \cite{ai6090216}. Grounded by these developments, recent research trends point toward the development of context-aware AI for threat analysis, incorporating graph neural networks, large language models, and explainable reasoning methods to support accurate and interpretable detection of multi-stage attacks, including those that target industrial and cyber–physical systems~\cite{9471816, info16121036}.

Against this backdrop, \textit{BRIDG-ICS} advances the state of the art by providing a contextual, AI-driven framework for multi-stage threat reasoning in Industry~5.0 environments. The framework integrates heterogeneous IT–OT–CPS knowledge and enriches it through NER–RE–NL2KG semantic reasoning, enabling dynamic expansion of the knowledge graph beyond what static models capture. It performs graph-based causal inference, reconstructs multi-hop attack paths, and supports probabilistic risk propagation, linking early indicators to operational impacts. By emphasizing explainability, analyst support, and proactive threat simulation, BRIDG-ICS offers a unified and interpretable capability that surpasses existing IT-only or OT-only approaches and strengthens advanced industrial cybersecurity.

\section{BRIDG-ICS: Knowledge Graphs for Intelligent Threats Analytics}

\begin{figure*}[!t]
    \centering
    \includegraphics[scale=.50]{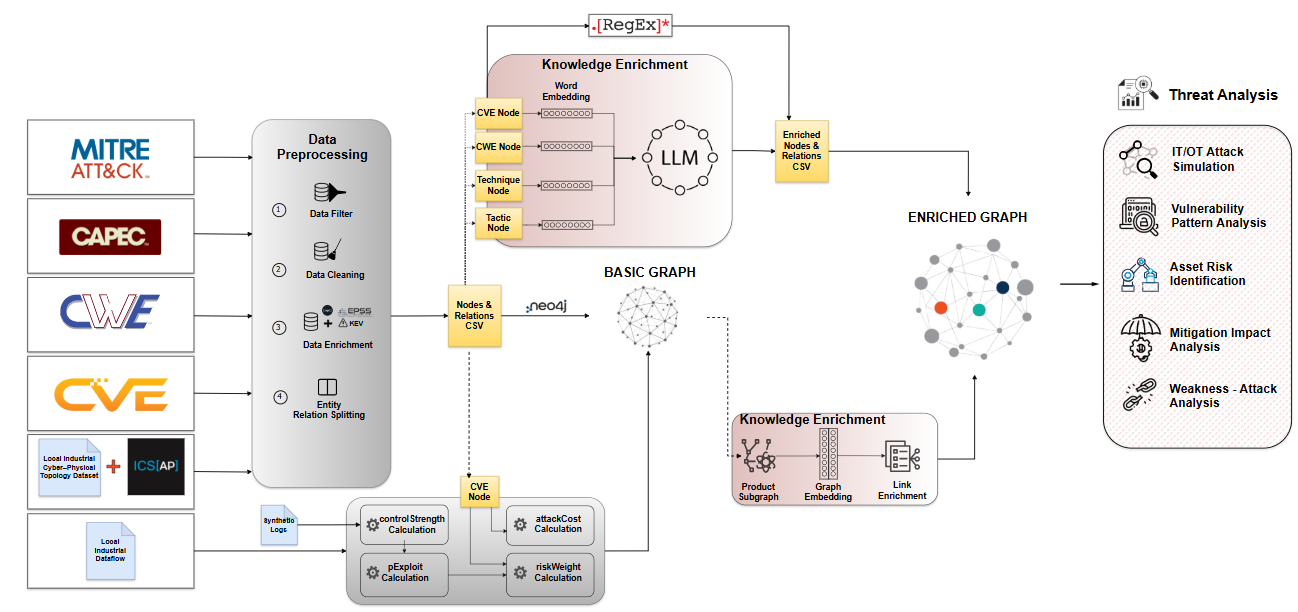}
    \caption{Overview of the BRIDG-ICS methodology.}
    \label{fig:approach}
\end{figure*}

This section introduces the \textit{BRIDG-ICS} framework (Figure~\ref{fig:approach}), which unifies ICS and cybersecurity data into a semantic knowledge graph. Extending  prior work ~\cite{ics_sec}, it integrates public sources including \textit{ICSA}~\cite{icsa}, \textit{CVE}~\cite{cve}, \textit{CPE}~\cite{cpe}, \textit{CAPEC}~\cite{capec}, \textit{CWE}~\cite{cwe}, and \textit{MITRE ATT\&CK}~\cite{mitre_ics}, covering the period 2002–2025. These heterogeneous CSV and XML datasets are transformed into contextualized knowledge instances, where nodes and relations represent assets, vulnerabilities, attack patterns, and adversarial behaviors. Through this unified representation and its AI-assisted enrichment mechanisms, BRIDG-ICS provides a foundation capable of addressing long-standing gaps in semantic integration, threat reasoning, and ICS-specific risk modelling.

In \textit{BRIDG-ICS}, the MITRE ATT\&CK framework serves as a structured ontology 
representing adversarial tactics and techniques across both IT and OT domains. Previous 
efforts attempted to map ICS advisories directly to ATT\&CK malware entries; however, this 
proved impractical in practice, as only a single advisory contained such a mapping. To 
avoid sparse or misleading connections, direct malware links are not incorporated. Instead, 
adversarial behavior is modeled through CVEs, CWEs, and CAPECs, which act as intermediary 
layers between ICS advisories and high-level ATT\&CK tactics and techniques. This 
semantics-driven design captures how attackers exploit documented weaknesses and enables 
scalable inference of multi-stage attack paths across IT/OT environments.

\begin{figure}[t]
\centering

% ==== Left image + caption (a) ====
\begin{minipage}{0.48\linewidth}
    \centering
    \includegraphics[width=\linewidth]{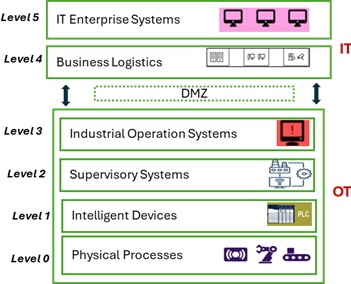}

    % Caption box of equal height
    \vspace{4pt}
    {\small\parbox[c][0.8cm][c]{\linewidth}{\centering \textbf{(a)} Overview of Local Dataset.}}
\end{minipage}
\hfill
% ==== Right image + caption (b) ====
\begin{minipage}{0.48\linewidth}
    \centering
    \vspace{30pt}
    \includegraphics[width=\linewidth]{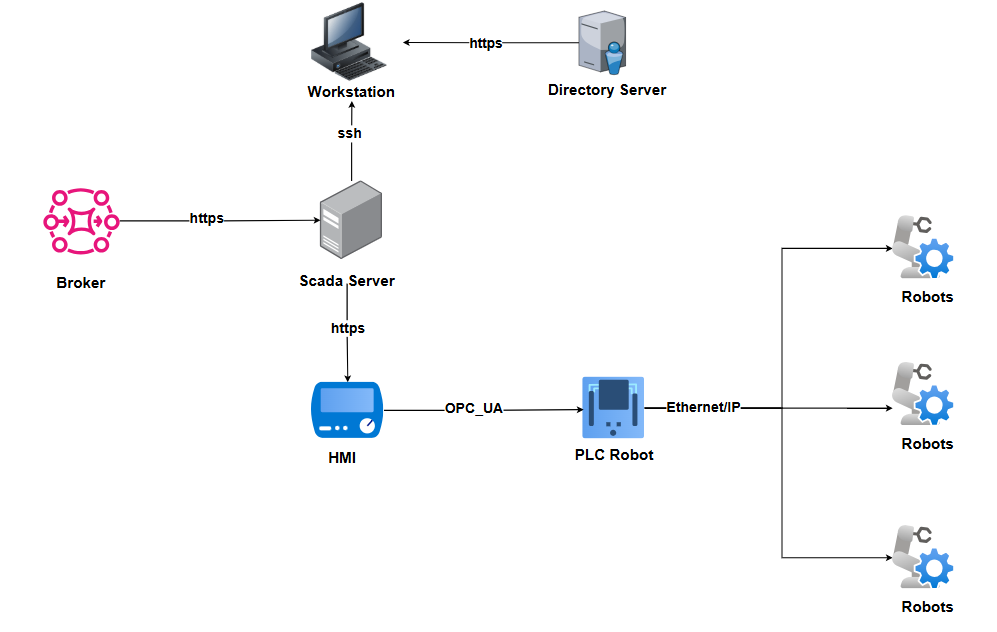}

    % SAME caption box height
    \vspace{4pt}
    {\small\parbox[c][0.8cm][c]{\linewidth}{\centering \textbf{(b)} Example of Local Dataflow.}}
\end{minipage}

\vspace{6pt}
\caption{Architecture and dataflow for the local dataset.}
\label{fig:local-dataflow}
\end{figure}

To ground the knowledge graph in realistic industrial environments, the proposed framework integrates 
a detailed industrial testbed directly into its ontology. The testbed follows the 
ISA-95/IEC~62264 Purdue architecture and provides rich metadata covering PLCs, 
SCADA/HMI systems, sensors, actuators, MES/ERP components, communication protocols, 
network zones, and deployment configurations. These elements are explicitly represented 
as interconnected nodes, supplying the cyber-physical context necessary for interpreting 
vulnerabilities and system behavior. The overall structure of this environment, along 
with examples of how local datasets are incorporated into the KG, is depicted 
in Figures~\ref{fig:local-dataflow}a and~\ref{fig:local-dataflow}b.

% \begin{center}
%   \includegraphics[width=0.6\linewidth]{dataflow.png}
%   \captionof{figure}{Example of Local Dataflow.}
%   \label{fig:dataflow}
% \end{center}

In addition to static asset modeling, the framework incorporates dynamic behavior through 
the inclusion of \emph{synthetic operational logs}. Using the CONTEXT dataset,
realistic OPC~UA and sensor telemetry traces were generated to reflect both baseline 
operations and security-enhanced configurations, including network segmentation, intrusion 
detection, and structured patch management\footnote{\href{https://github.com/ahmadspm/Industry-5.0--Intelligent-Threat-Analytics-KGs-and-LLMs/tree/main/Phase1-KG/Synthetic_Data}{Synthetic Data Repository}}. 
This augmentation ensures that BRIDG-ICS captures not only the structural layout of 
industrial systems but also the behavioral patterns that influence threat dynamics.
.

\subsection{Ontology Design}

Following from \cite{ics_sec} and \cite{shen2020}, we designed an ontology that unifies industrial operations and cybersecurity within a single semantic framework. It captures interactions among physical components, communication processes, and digital threats, forming a basis for attack simulation, risk assessment, and resilience analysis. As shown in Figure~\ref{fig:ontology}, the BRIDG-ICS ontology comprises five linked sub-ontologies: (i) ICS, (ii) CVE, (iii) CWE, (iv) CAPEC, and (v) MITRE ATT\&CK. These sub-ontologies are arranged hierarchically, connecting low-level industrial assets to higher-level adversarial behaviors and tactics.

\begin{center}
  \includegraphics[width=0.75\linewidth]{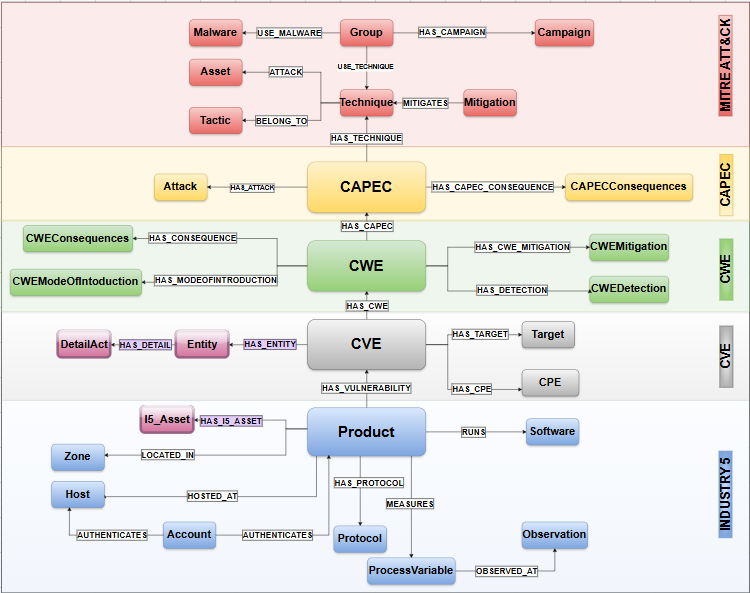}
  \captionof{figure}{BRIDG-ICS Ontology.}
  \label{fig:ontology}
\end{center}
\vspace{0.75em} 

\textbf{\textit{(i) ICS Ontology}} models smart manufacturing environments and defines core classes such as \textit{Product}, \textit{Software}, \textit{Component}, \textit{Protocol}, \textit{Zone}, \textit{Account}, \textit{ProcessVariable}, and \textit{Observation}.

\textbf{\textit{(ii) CVE Ontology}} models vulnerability information from CVE records, unifying \textit{EPSS}, \textit{KEV}, and \textit{CPE} metadata into a single \textit{Vulnerability} entity. It also incorporates vendor statements from NVD (e.g., \textit{``Red Hat Enterprise Linux~5 is not vulnerable to this issue as it contains a backported patch''}) to capture mitigations, affected versions, and vendor responses. To our knowledge, such contextual data have not appeared in prior knowledge graphs, making this a novel contribution that improves interpretability and precision in vulnerability reasoning.

\textbf{\textit{(iii) CWE Ontology}} models software and hardware weaknesses through entities such as \textit{CWE}, \textit{Mitigation}, \textit{Detection}, \textit{ModeOfIntroduction}, and \textit{Consequence}. Figure~\ref{fig:semantic} illustrates the semantic structure of this sub-ontology, including the introduced \texttt{COMMUNICATES\_WITH} relation, which represents dataflow interactions observed in the industrial testbed.

\textbf{\textit{(iv) CAPEC Ontology}} captures adversarial attack patterns used to exploit software or hardware weaknesses through entities such as \textit{CAPEC}, \textit{Consequence}, and \textit{Attack}.

\textbf{\textit{(v) miter ATT\&CK Ontology}} models adversarial tactics, techniques, and procedures (TTPs), encompassing entities such as \textit{Technique}, \textit{Malware}, \textit{Mitigation}, \textit{Tactic}, \textit{Group}, and \textit{Asset}.

\vspace{-0.55em} 
\begin{center}
  \includegraphics[width=0.55\linewidth]{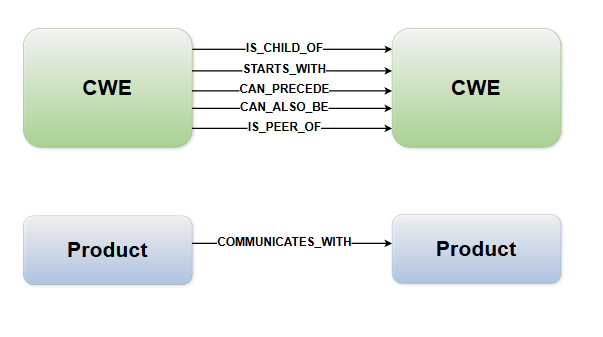}
  \vspace{-0.75em} 
  \captionof{figure}{Interrelationships between CWE and Product.}
  \label{fig:semantic}
\end{center}

\subsection{Basic Knowledge Graph Development}

The first stage of BRIDG-ICS involves constructing the foundational knowledge graph by 
integrating heterogeneous cybersecurity datasets with operational data from the industrial 
testbed. This process consists of three steps: preprocessing public datasets, incorporating 
local industrial context, and generating the final structured graph representation.

\textbf{Preprocessing.} To ensure consistency across all sources, CVE entries marked as 
\textit{REJECTED} or \textit{RESOLVED} were removed, and textual fields were standardized by 
stripping non-essential characters. Product names from ICSA advisories were retained in 
their original form to preserve natural variability, and each product was linked to its 
corresponding CPE entry, enabling CVE vulnerabilities to be associated with generalized 
vendor and product identifiers. The CWE and CAPEC datasets, originally provided in XML, were 
parsed and converted into CSV for uniform integration into the ontology.

\sloppy{
\textbf{Local Industry Testbed.} Operational data from the local industrial testbed were 
integrated into the ontology to complement the standardized public datasets. Industrial 
product entries were aligned with ICSA records, ensuring consistent mappings between 
\textit{Product} components and associated vulnerabilities. Communication dependencies 
among components were derived using the \textit{COMMUNICATES\_WITH} relation, informed by 
observed protocols and data flows. Testbed logs were also used to generate quantitative 
metrics for both \textit{normal operation} and a \textit{secured} configuration, providing 
behavioral context for analyzing system dynamics and attack-path propagation.}

\textbf{Integration Output.} The final integration produced two structured outputs: a 
\texttt{node.csv} and a \texttt{relation.csv}, both imported into the knowledge engine for 
graph instantiation. Entities from CVE, CWE, CAPEC, CPE, and ATT\&CK were mapped to 
ontology classes to ensure semantic consistency and interoperability across domains. 
Relationships followed a hierarchical structure:
\[
\textit{ICS} \rightarrow \textit{CVE} \rightarrow \textit{CWE} \rightarrow \textit{CAPEC} 
\rightarrow \textit{MITRE ATT\&CK},
\]
providing a unified representation of how industrial products connect to vulnerabilities, 
weaknesses, exploitation patterns, and adversarial techniques within the BRIDG-ICS 
Knowledge Graph.

\subsection{Risk Modelling and Attack Path Formulation}
\sloppy{
Risk modelling in \textit{BRIDG-ICS} is built upon the interactions between industrial 
\textit{Products}, captured through the \texttt{COMMUNICATES\_WITH} and 
\texttt{CONTROLLED\_COMMUNICATES\_WITH} relations. These relations are derived from dataflow analysis, which identifies protocol-level connections such as \texttt{Modbus/TCP}, \texttt{OPC\_UA}, and \texttt{PROFINET}, along with their corresponding endpoints. Each relation is enriched with quantitative attributes including \textit{riskWeight}, \textit{pExploit}, and \textit{attackCost}, derived from system activity logs that capture communication patterns, authentication events, and configuration states. This parametrization enables the \textbf{BRIDG-ICS} framework to model attack propagation across interconnected components with greater accuracy.}

\subsubsection{COMMUNICATES\_WITH}
\sloppy{
The \texttt{COMMUNICATES\_WITH} relation captures the residual risk between 
two products \(u\) and \(v\) during normal operations. Each connection links 
IT and OT components and is associated with the attribute set  
\( A = \{\textit{riskWeight}, \textit{pExploit}, \textit{attackCost}\} \), 
which characterizes the security posture of the communication link. 
}
To express the level of protection on each path, a composite 
control-effectiveness metric, \texttt{controlStrength}, is defined as:
\begin{equation}
\texttt{controlStrength}(u,v) = a \times c \times e \times h,
\tag{3}
\label{eq:controlStrength}
\end{equation}
where \(a\), \(c\), \(e\), and \(h\) represent accessibility, configuration hygiene, 
exploitability resistance, and residual hardening effectiveness.  
These values are derived from operational logs such as authentication patterns, 
misconfiguration rates, failed write attempts, and certificate usage, yielding 
a quantitative estimate of the defensive strength of each communication link.  
Lower \texttt{controlStrength} values indicate weaker protection and higher 
susceptibility to compromise.

Based on this defensive estimation, the likelihood of exploitation is obtained 
by attenuating the Exploit Prediction Scoring System (EPSS) value with the 
observed control strength:
\begin{equation}
p_{\textit{Exploit}}(u,v) = \texttt{EPSS}(u,v)\,[1 - \texttt{controlStrength}(u,v)]
\tag{4}
\label{eq:pExploit}
\end{equation}

In this formulation, \(\texttt{EPSS}(u,v)\) denotes the exploitation probability associated with the 
corresponding vulnerability. When multiple CVEs are linked to a single communication 
path, the overall likelihood is computed using the aggregation function  
\(1 - \prod_i (1 - e_i)\), which represents the probability that at least one of the 
individual exploits succeeds.

To represent adversarial effort, an \texttt{attackCost} parameter quantifies the 
complexity and severity of potential exploits:
\begin{equation}
\texttt{attackCost}(u,v) = \text{Base} + f_{AC} + f_{AV} + \text{EPSS},
\tag{5}
\label{eq:attackCost}
\end{equation}
where \text{Base} is derived from the CVSS vector, and  
\(f_{AC}\) and \(f_{AV}\) encode access complexity and attack vector categories.  
Higher \texttt{attackCost} values suggest greater resource requirements or 
technical difficulty for an attacker.

The overall residual risk for each connection is then expressed through the 
\texttt{riskWeight} attribute:
\begin{equation}
\texttt{riskWeight}(u,v) = p_{\textit{Exploit}}(u,v) \times \frac{\texttt{criticality}(v)}{10},
\tag{6}
\label{eq:riskWeight}
\end{equation}
combining the probability of exploitation with the importance of the target asset.  
Within the graph model \(G = (V, E, \phi, \psi)\), these quantitative attributes  
\(A = \{\textit{riskWeight}, p_{\textit{Exploit}}, \textit{attackCost}\}\)  
are assigned to edges through the mapping  
\(\psi : E \rightarrow T_E \times A\), enabling reasoning across both structural 
and quantitative dimensions of cyber–physical relations.

\paragraph{\textbf{Attack Path and Exposure Estimation.}}
Each directed edge \(E(u,v)\) incorporates the parameters defined in 
Eqs.~(\ref{eq:controlStrength})–(\ref{eq:riskWeight}), making it possible to model 
how attacks may propagate across interconnected components.  
The likelihood of a multi-hop attack along a path  
\(P = \langle v_1, v_2, \ldots, v_k \rangle\)  
is computed as:
\begin{equation}
p(P) = \prod_{i=1}^{k-1} p_{\textit{Exploit}}(v_i, v_{i+1}),
\tag{7}
\label{eq:attack_path_prob}
\end{equation}
representing the cumulative probability that an adversary successfully advances 
through each step. Larger \(p(P)\) values correspond to more plausible routes within the system.

Node-level exposure is then calculated as the weighted sum of incoming risks:
\begin{equation}
\textit{Exposure}(v) = \sum_{(u,v) \in E} \textit{riskWeight}(u,v),
\tag{8}
\label{eq:node_exposure}
\end{equation}
which reflects the overall exploitation pressure on each asset.  
This metric highlights critical components and supports resilience analysis 
through centrality and modularity evaluations.  
Together, Eqs.~(\ref{eq:attack_path_prob})–(\ref{eq:node_exposure}) establish a 
coherent basis for estimating attack feasibility and systemic vulnerability 
across cyber–physical networks.

\subsubsection{CONTROLLED\_COMMUNICATES\_WITH}

The \texttt{CONTROLLED\_COMMUNICATES\_WITH} relation is structurally identical to 
\texttt{COMMUNICATES\_WITH}, but it represents communication links in which the 
associated products operate under security controls aligned with NIST SP 80053 
and IEC 62443. The presence of these controls influences the operational logs, 
including authentication events, protocol utilization, and configuration states. 
Consequently, the resulting communication behavior exhibits stronger security 
properties and a lower level of residual risk compared to the uncontrolled case.

Operational logs updated to reflect the applied controls, including network 
segmentation, access restrictions, and patch management, are subsequently used to 
recompute the parameters defined in Eqs.~(\ref{eq:controlStrength})--(\ref{eq:node_exposure}). 
For each communication pair \((u,v)\), the recalculated values typically indicate an 
increase in \texttt{controlStrength} and a reduction in \(p_{\textit{Exploit}}\), which in 
turn lowers the corresponding \texttt{riskWeight}. As a result, both attack path 
probabilities \(p(P)\) and node exposures \(\textit{Exposure}(v)\) decrease, reflecting the 
quantifiable impact of the implemented security measures. Links with 
\(\texttt{riskWeight}(u,v) < 0.05\) are treated as non-exploitable and are removed from 
subsequent attack path simulations. This controlled comparison provides a means to 
assess and visualize changes in systemic exposure before and after the application 
of defensive controls. A summary of the implemented controls and their effects is 
presented in Table~\ref{tab:implemented_controls}.

\begin{table}[htbp]
\centering
\caption{Mapping of Implemented Controls to BRIDG-ICS Control Factors}
\label{tab:implemented_controls}
\renewcommand{\arraystretch}{1.1} % Adjust vertical spacing
\setlength{\tabcolsep}{4pt} % Adjust horizontal padding

\begin{tabularx}{\columnwidth}{|p{3cm}|X|}
\hline
\textbf{Control Type} & \textbf{Description / Implementation Effect} \\
\hline\hline

\textbf{Network Segmentation} &
Implements Purdue-based isolation between IT and OT zones, limiting lateral movement and enforcing access boundaries through VLANs and firewalls. \\ \hline

\textbf{Patch Management} &
Applies security updates to firmware, OS, and applications, mitigating known CVEs and reducing residual exploitability. \\ \hline

\textbf{Intrusion Detection (IDS)} &
Deploys anomaly- and signature-based detection within OT network segments to detect failed writes, unauthorized access, and malicious traffic. \\ \hline

\textbf{Access Control \& Authentication} &
Uses certificate-based authentication, least privilege access, and enforced user roles to prevent insecure or anonymous sessions. \\ \hline

\textbf{Configuration Hardening} &
Removes default credentials, enforces secure configurations, disables unused services, and applies compliance baselines. \\ \hline

\end{tabularx}
\end{table}

The recalculated metrics under both controlled and uncontrolled conditions make it 
possible to examine how defensive measures reshape the security posture of 
individual communication links and the broader system. These parameters, defined in 
Eqs.~(\ref{eq:controlStrength})--(\ref{eq:node_exposure}), form the core analytical 
basis for resilience evaluation in the framework. The relations 
\texttt{COMMUNICATES\_WITH} and \texttt{CONTROLLED\_COMMUNICATES\_WITH} encode the 
relationship between exploitation likelihood and defensive effectiveness, allowing 
risk to propagate through interconnected assets.

These formulations are operationalized through graph analytical techniques, including 
PageRank, betweenness centrality, and Louvain modularity. In this analysis, 
\texttt{riskWeight}, \texttt{attackCost}, and \(p_{\textit{Exploit}}\) serve as 
edge weights to identify critical nodes, high probability attack paths, and the 
overall impact of mitigation strategies on systemic resilience.

\subsection{Knowledge Enrichment}

Knowledge enrichment in BRIDG-ICS addresses the semantic gaps, incomplete mappings, and inconsistencies inherent in publicly available cybersecurity datasets. Many CVE entries lack corresponding CWE classifications, ATT\&CK technique associations, or contextual indicators relevant to industrial cyber–physical systems. Furthermore, CVE descriptions frequently embed operational semantics, such as device behaviors, exploitation prerequisites, and misconfiguration patterns, that are not represented in structured metadata. To address these limitations, BRIDG-ICS uses contextual embeddings from security-domain transformer models (LLMs), including \textit{SecureBERT} and \textit{CySecBERT}, to infer missing associations, tighten semantic proximity among related entities, and improve overall graph coherence. This embedding-based augmentation uncovers latent relationships and improves KG completeness by reinforcing structurally and semantically plausible links. For details, see the knowledge enrichment step in the proposed approach in \ref{fig:approach}.

Beyond embeddings, BRIDG-ICS employs a suite of \textit{LLM-assisted enrichment techniques} to extract threat semantics from unstructured CTI reports, ICS advisories, and vulnerability narratives. Three key processes underpin this enrichment pipeline:

\begin{itemize}
\item \textbf{Named Entity Recognition (NER):} Using \textit{SecureBERT}, the system identifies domain-specific artifacts such as tools, file objects, network paths, and configuration parameters. These artifacts are represented as typed nodes, enriching the granularity of exploit- and behavior-level knowledge.
\item \textbf{Relation Extraction (RE):} Semantic relations, including \textit{uses}, \textit{modifications}, \textit{targets}, and \textit{effects}, are extracted via the \textit{REBEL} sequence-to-sequence model, enabling the KG to represent how vulnerabilities manifest in realistic attack scenarios.

\item \textbf{Natural Language–to–Knowledge Graph (NL2KG):} LLM-driven transformation pipelines convert natural language threat descriptions into ontology-aligned triples, expanding cross-domain (IT–OT–CPS) associations that are not explicitly encoded in structured datasets.
\end{itemize}

For relation inference, BRIDG-ICS uses transformer-based classifiers trained on domain-specific embeddings to reconstruct missing CVE→CWE and CVE→ATT\&CK mappings. The resulting enriched KG provides a high-fidelity substrate for graph-based threat analytics, probabilistic attack-path modeling, and resilience assessment. Moreover, when used as grounding knowledge for retrieval-augmented LLM applications, the enriched KG constrains generative outputs to verified industrial cybersecurity facts, reducing unsupported inferences and improving the interpretability and reliability of AI-assisted threat reasoning.
\section{Result and Analysis}
This section presents the outcomes of the experiments derived from the BRIDG-ICS framework.  We evaluate the effectiveness of the knowledge enrichment procedures, analyze inferred relations and prediction performance, and examine how the enriched knowledge graph supports threat propagation assessment and cyber–physical 
resilience analysis.

\subsection{Enrichment Experiment Setup}

All experiments in this section were conducted to support knowledge enrichment
within the BRIDG-ICS Knowledge Graph, to expand its semantic coverage and
inferring relations missing from existing public cybersecurity datasets.

The enrichment workflows were executed on a workstation with an NVIDIA RTX~6000 Ada GPU (48 GB VRAM), CUDA~13.0, 64 GB RAM, and a multi-core CPU. Transformer models were implemented in \texttt{PyTorch~2.8.0} using \texttt{HuggingFace Transformers}, and their embeddings and inferred relations were integrated into the Neo4j~5.x \textit{KG engine} used throughout this work for both graph construction and inference. Graph reasoning was performed using the GDS~2.x library, including FastRP embeddings, KNN similarity, $k$-shortest paths, PageRank, and Betweenness, providing the computational foundation for the enrichment experiments described below.

\medskip
\noindent\textbf{Node Enrichment.}
To capture contextual information not present in structured CVE fields, semantic features
were extracted directly from CVE descriptions. Named Entity Recognition was performed using
\textit{SecureBERT}, followed by Relation Extraction using \textit{REBEL}. Extracted
elements, such as \textit{TOOL}, \textit{FILE}, and \textit{URL}, and their associated
\textit{DetailAct} relations were normalized and incorporated as typed \textit{Entity}
nodes within the ontology. This process groups CVEs referencing similar artifacts and
enables downstream models to recognize shared exploitation behaviors.
\newline 
\textbf{Relation Enrichment.}
A complementary set of experiments focused on inferring missing semantic links. Many CVEs lack mappings to CWE weakness classes or ATT\&CK techniques, which constrain higher-level analysis. To infer these associations, CVE text was embedded using \textit{SecureBERT} and evaluated using three classification strategies: K-Nearest Neighbors (KNN) over FastRP embeddings, a fine-tuned Multi-Layer Perceptron (MLP), and a BERT-based classifier. Hyperparameters were kept consistent across models to enable comparable evaluation (Table~\ref{tab:hyperparams}). For the \texttt{HAS\_POSSIBLE\_CWE} relation, the BERT classifier with \textit{CySecBERT} embeddings achieved the best validation accuracy (66.47\%), as shown in Table~\ref{tab:relation_enrichment}.

Inference of \texttt{HAS\_POSSIBLE\_TECHNIQUE} links followed a similar setup, testing
contextual embeddings from \textit{AttackBERT} and \textit{SecRoBERTa} with the same
classification models over five training epochs. The strongest results were obtained using
\textit{SecRoBERTa} with the CyberBERT classifier, which reached 98.70\% accuracy
(Table~\ref{tab:tech_results}), demonstrating the advantages of security domain
pretraining.

\begin{table}[h]
\centering
\caption{Training Hyperparameters for Enrichment Experiments}
\label{tab:hyperparams}
\begin{tabular}{lc}
\hline
\textbf{Parameter} & \textbf{Value} \\
\hline
Learning Rate & \(2 \times 10^{-5}\) \\
Batch Size & 128 \\
Optimizer & Adam \\
Loss Function & Cross-Entropy Loss \\
Epochs & 20 (CWE), 5 (Technique) \\
Validation Split & 20\% \\
\hline
\end{tabular}
\end{table}

\begin{table}[h]
\centering
\caption{Validation Accuracy for \texttt{HAS\_POSSIBLE\_CWE} Enrichment (\%)}
\label{tab:relation_enrichment}
\begin{tabular}{lcc}
\hline
\textbf{Model} & \textbf{CyBERT} & \textbf{CySecBERT} \\
\hline
KNN & 64.55 & 64.70 \\
MLP & 63.18 & 64.20 \\
BERT Classifier & 63.63 & \textbf{66.47} \\
\hline
\end{tabular}
\end{table}

\textbf{Additional Relation Types.}
Further enrichment included the generation of potential communication links
(\texttt{HAS\_POSSIBLE\_COMMUNICATION}) by connecting each product to its five most
similar nodes using KNN over FastRP embeddings. This reveals plausible but unobserved
interactions that may exist in real industrial environments. Tactic-level predictions
(\texttt{SUGGESTED\_TACTIC}) were produced using a pretrained BERT
model\footnote{\url{https://huggingface.co/sarahwei/MITRE-v16-tactic-bert-case-based}}
with 0.9887 accuracy.

\begin{table}[h]
\centering
\caption{Validation Accuracy for \texttt{HAS\_POSSIBLE\_TECHNIQUE} Enrichment (\%)}
\label{tab:tech_results}
\begin{tabular}{lcc}
\hline
\textbf{Model} & \textbf{AttackBERT} & \textbf{SecRoBERTa} \\
\hline
KNN & 10.20 & 51.00 \\
MLP & 14.73 & 60.63 \\
CyberBERT Classifier & 61.04 & \textbf{98.70} \\
\hline
\end{tabular}
\end{table}

\subsection{Threat Analysis and Propagation Analysis}
To understand the operational value of the enriched BRIDG-ICS graph, a series of 
multi-stage attack simulations is performed to illustrate how adversaries may traverse 
interconnected IT–OT environments. The knowledge graph provides a foundation for 
\textit{threat analysis} by exposing previously hidden dependencies, identifying 
high-influence assets, and quantifying propagation behavior across industrial systems. 
The impact of enrichment and security controls is assessed with respect to attack-path 
length, asset reachability, and overall system resilience. The evaluation encompasses 15 
smart manufacturing scenarios covering three intrusion phases: attack surface expansion, 
lateral movement, and crown-jewel impact. Detailed descriptions of all scenarios are 
provided in Appendix~\ref{ap:att-scenario}.

\subsubsection*{Evaluation Framework}

Attack propagation was assessed using the \texttt{COMMUNICATES\_WITH} relation, which 
captures logical or protocol-level connectivity between products. For each scenario, the 
analysis measured the number of hops an adversary could traverse from a compromised source 
to a designated target asset. The evaluation considered three configurations:

\vspace{-4pt}
\begin{itemize}
    \item \textbf{Original:} the baseline topology containing only direct communication 
    links or observed dataflow;
    \item \textbf{Enriched:} the baseline graph augmented with inferred 
    \texttt{HAS\_POSSIBLE\_COMMUNICATION} relations generated through FastRP–KNN 
    similarity, revealing latent or unobserved dependencies;
    \item \textbf{Controlled:} the enriched graph with NIST-aligned mitigations applied, 
    where edges with \texttt{riskWeight} below 0.05 are treated as non-exploitable.
\end{itemize}
\vspace{-4pt}

Propagation behavior across these configurations was examined using the Yen \(k\)-shortest 
paths algorithm (\texttt{gds.yenStream}) with \(k = 20\). Edges were treated as undirected, 
and \texttt{riskWeight} served as the cumulative cost metric governing path selection.

\begin{figure*}[t]
\centering

% --- Left Image (a) ---
\begin{minipage}{0.47\textwidth}
    \centering
    \includegraphics[width=\linewidth]{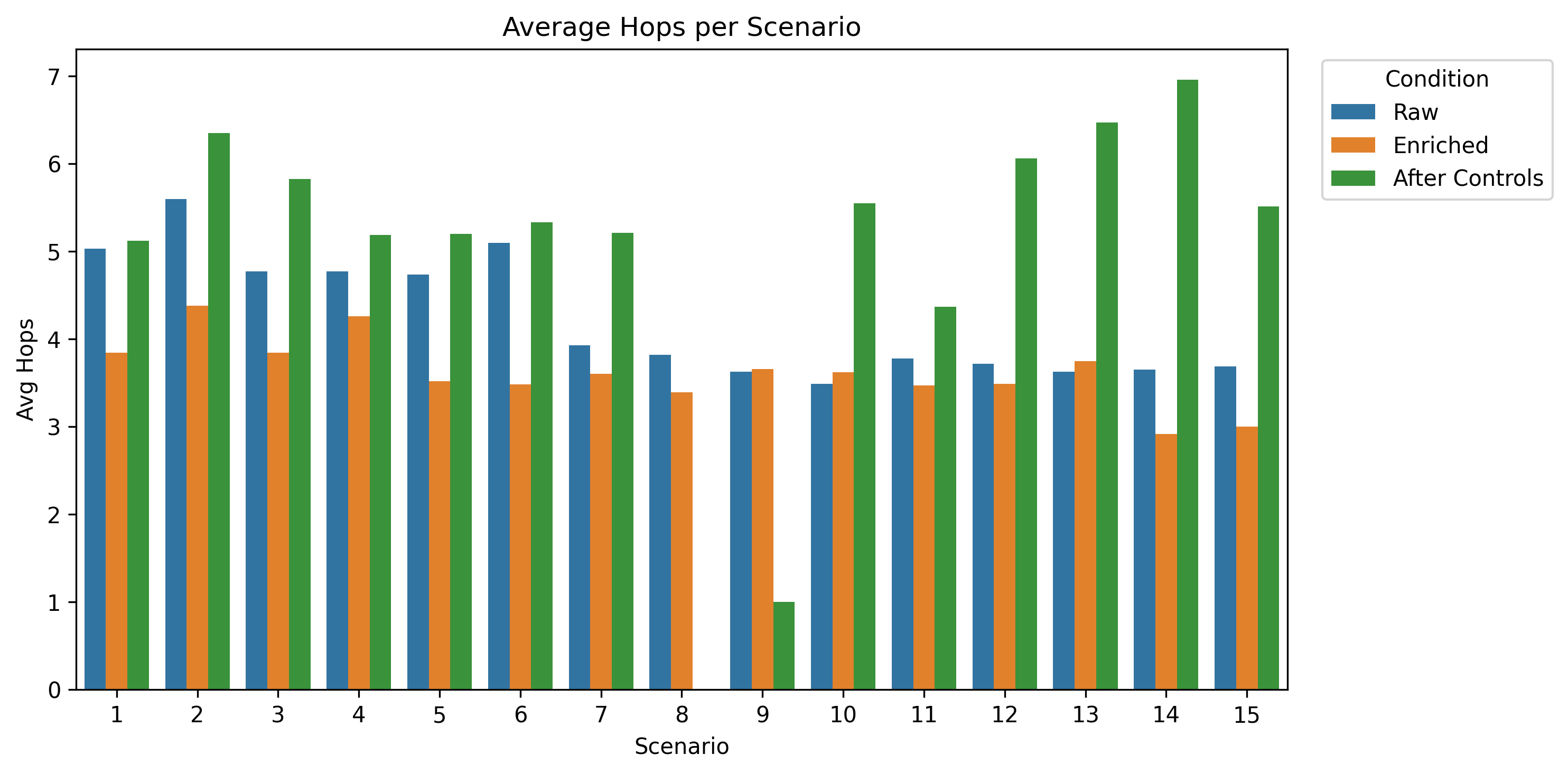}
    \textbf{(a)} Average hop count across 15 scenarios.
\end{minipage}
\hfill
% --- Right Image (b) ---
\begin{minipage}{0.47\textwidth}
    \centering
    \includegraphics[width=\linewidth]{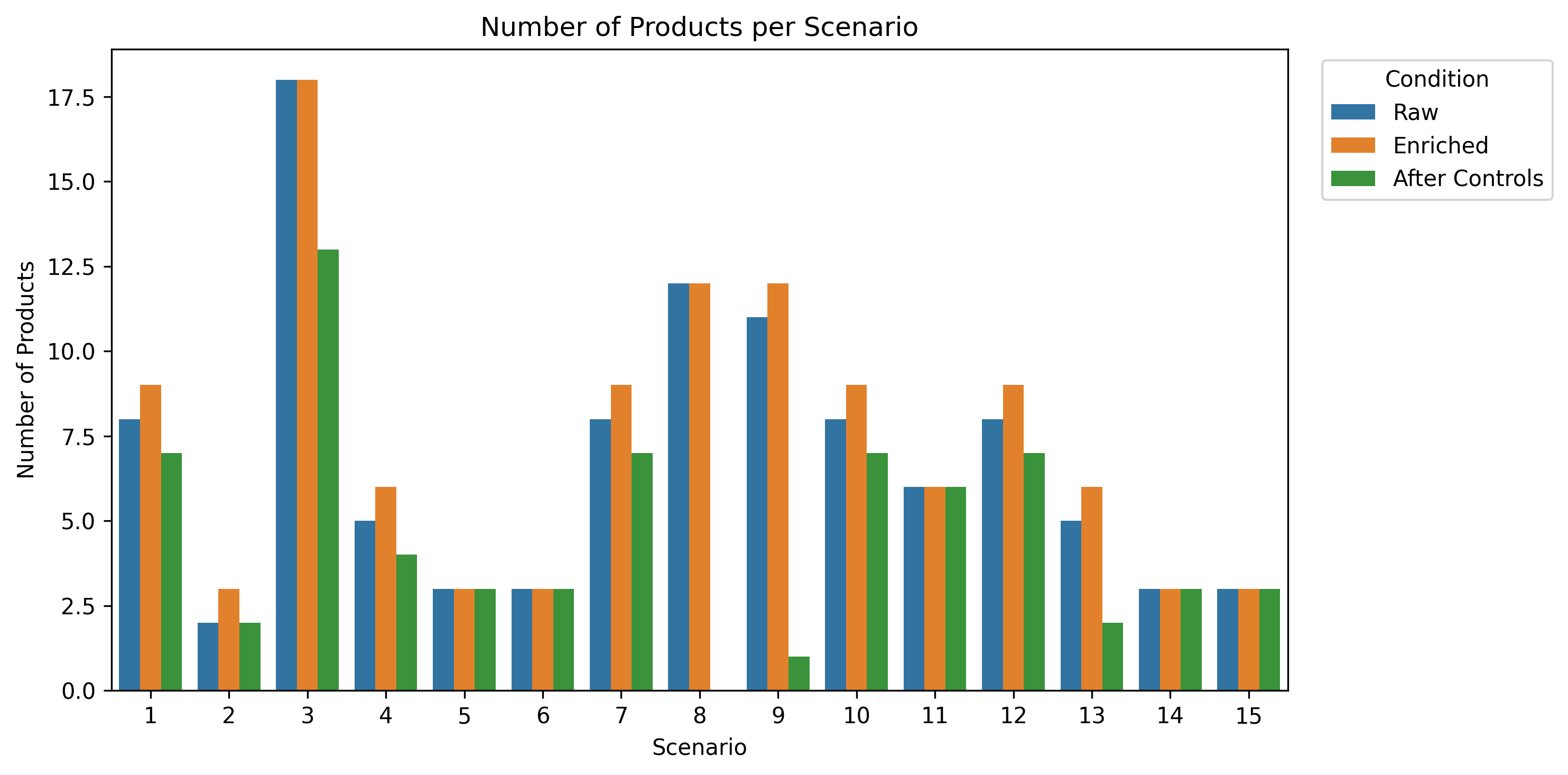}
    \textbf{(b)} Number of reachable products across 15 scenarios.
\end{minipage}
\vspace{8pt}
\caption{Comparison of propagation metrics across 15 simulated attack scenarios for the Original, Enriched, and Controlled configurations.}
\label{fig:overall_propagation}

\end{figure*}

\subsubsection*{Overall Propagation Results}
A comparison of attack-propagation behavior under the raw, enriched, and controlled 
configurations is presented in Figures~\ref{fig:overall_propagation}a 
and~\ref{fig:overall_propagation}b, covering fifteen representative scenarios.
In Figure~\ref{fig:overall_propagation}a, the \textbf{enriched} configuration consistently reduces the average number of hops required for an attacker to reach a target—typically by 20–40\% compared to the baseline. 
This reduction results from inferred \texttt{HAS\_POSSIBLE\_COMMUNICATION} links, which reveal previously unobserved pathways between devices and expose latent attack routes. 
In contrast, the \textbf{controlled} configuration increases average path lengths by 25–50\%, demonstrating the segmentation effect of applied mitigations that restrict lateral movement.

% Figure 8 — Attack-path length distribution (boxplot)
\begin{center}
  \includegraphics[scale=.46]{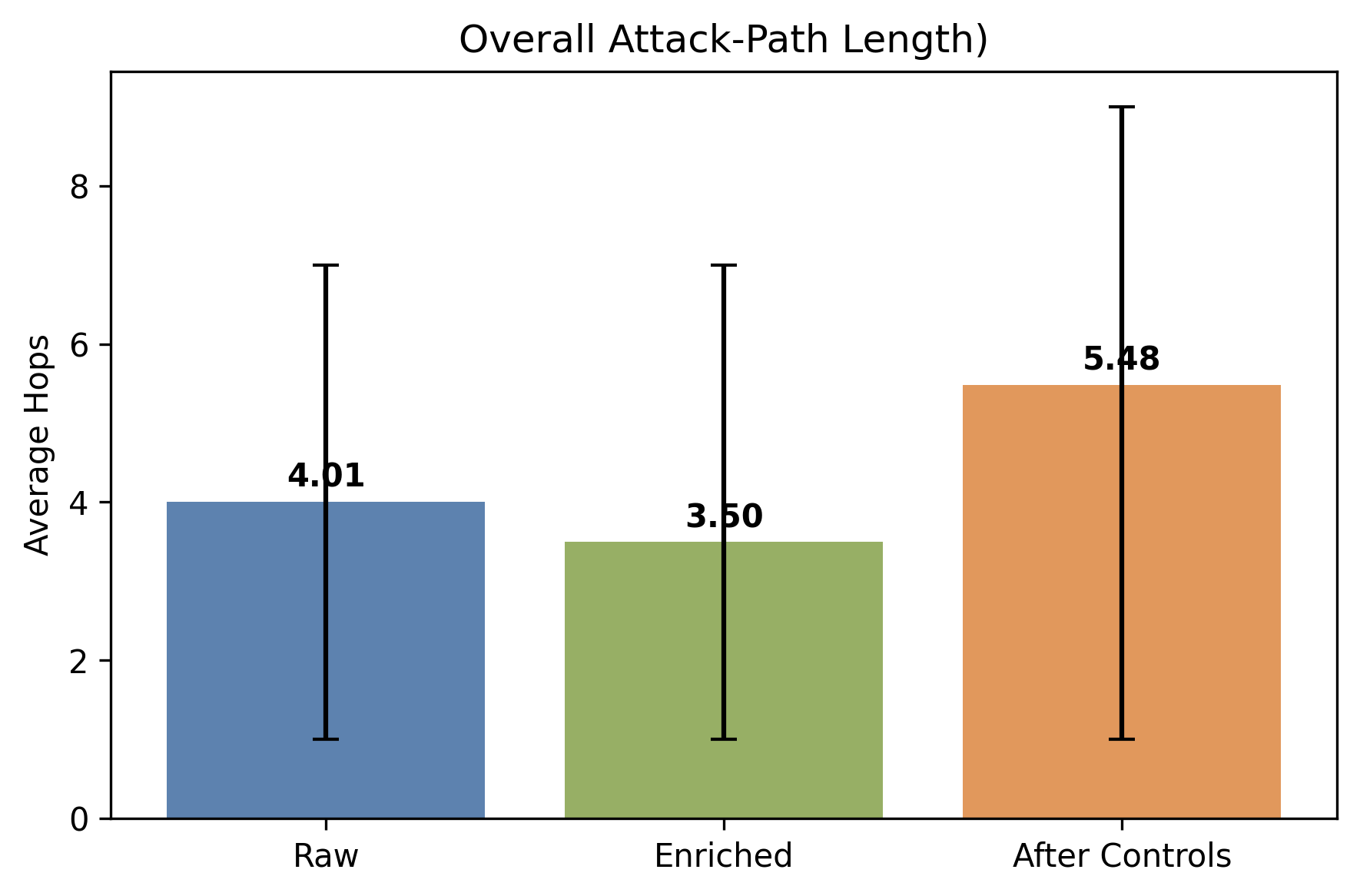}
  \captionof{figure}{Mean attack-path length with 95\% confidence intervals across the \textit{Raw}, \textit{Enriched}, and \textit{After Controls} knowledge-graph configurations.}
  \label{fig:mean_hops_ci}
\end{center}

Figure~\ref{fig:overall_propagation}b further supports these results by showing the change in the \textbf{number of affected products (reachable assets)} under each configuration. 
In most scenarios, enrichment increases the number of reachable products—from fewer than five to more than twelve—illustrating how inferred relations expand the visible attack surface. 
Conversely, the controlled configuration sharply reduces reachable assets to two to four in several cases, confirming that the implemented safeguards effectively limit propagation even in densely connected networks. 
Notably, scenarios~3, 8, and~12 exhibit the strongest contrasts between enriched and controlled conditions, reflecting the efficacy of segmentation and access restrictions.

Further supporting this observation, Figure~\ref{fig:mean_hops_ci} presents the \textit{average attack-path length} with 95\% confidence intervals across the \textit{Raw}, \textit{Enriched}, and \textit{After Controls} configurations. 
The enriched configuration slightly reduces the mean path length (\(\approx 3.5\)~hops) 
relative to the raw graph (\(\approx 4.0\)~hops), showing that link-prediction 
enrichment enhances propagation visibility. After applying NIST\textasciitilde SP\textasciitilde 800--53 
and IEC\textasciitilde 62443 controls, the mean path length increases to 
\(\approx 5.5\)~hops, indicating effective segmentation and hardening across 
IT--OT boundaries.

The non-overlapping confidence intervals demonstrate that BRIDG-ICS enrichment and mitigation translate semantic improvements in the knowledge graph into measurable reductions in adversarial reachability.

\begin{center}
  \includegraphics[scale=.46]{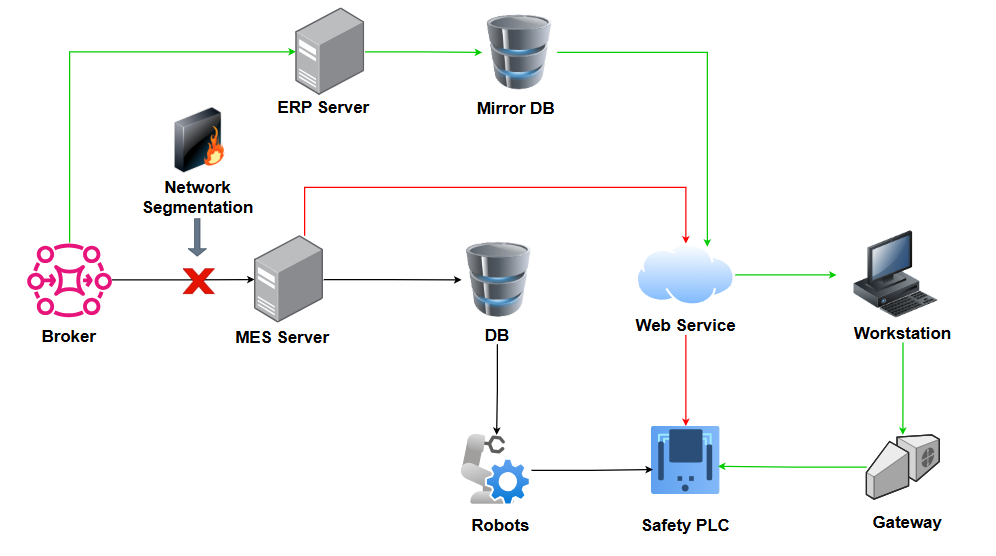}
  \captionof{figure}{Example Scenario From MQTT Broker to PLC}
  \label{fig:scenario1}
\end{center}

\subsubsection*{Representative Scenario: MQTT $\rightarrow$ PLC}
One of the scenario examples is a misconfigured MQTT broker in the DMZ that becomes the initial adversarial entry point. As shown in Table~\ref{tab:scenario1}, enrichment reduces the minimum attack path to a single hop and expands reachability to nine PLCs, exposing previously hidden lateral routes. When control mechanisms are applied, reachability decreases to seven PLCs and the average path length increases, indicating effective segmentation and reduced propagation potential within the control network.

\begin{table*}[t]
\centering
\caption{Hop comparison from MQTT Broker to PLC}
\label{tab:scenario1}

\begin{tabular}{l l l c c c c}
\toprule
Source & Target & Dataset & Avg. & Min & Max & Affected \\
\midrule
MQTT Broker & PLC & Original   & 5.03 & 4 & 6 & 8 \\
MQTT Broker & PLC & Enriched   & 3.84 & 1 & 6 & 9 \\
\textbf{MQTT Broker} & \textbf{PLC} & \textbf{Controlled} &
\textbf{5.12} & \textbf{4} & \textbf{5} & \textbf{7} \\
\bottomrule
\end{tabular}

\end{table*}

\begin{figure*}[t]
\centering

% ---------- Left image (a) ----------
\begin{minipage}{0.48\textwidth}
    \centering
    \includegraphics[width=0.95\linewidth]{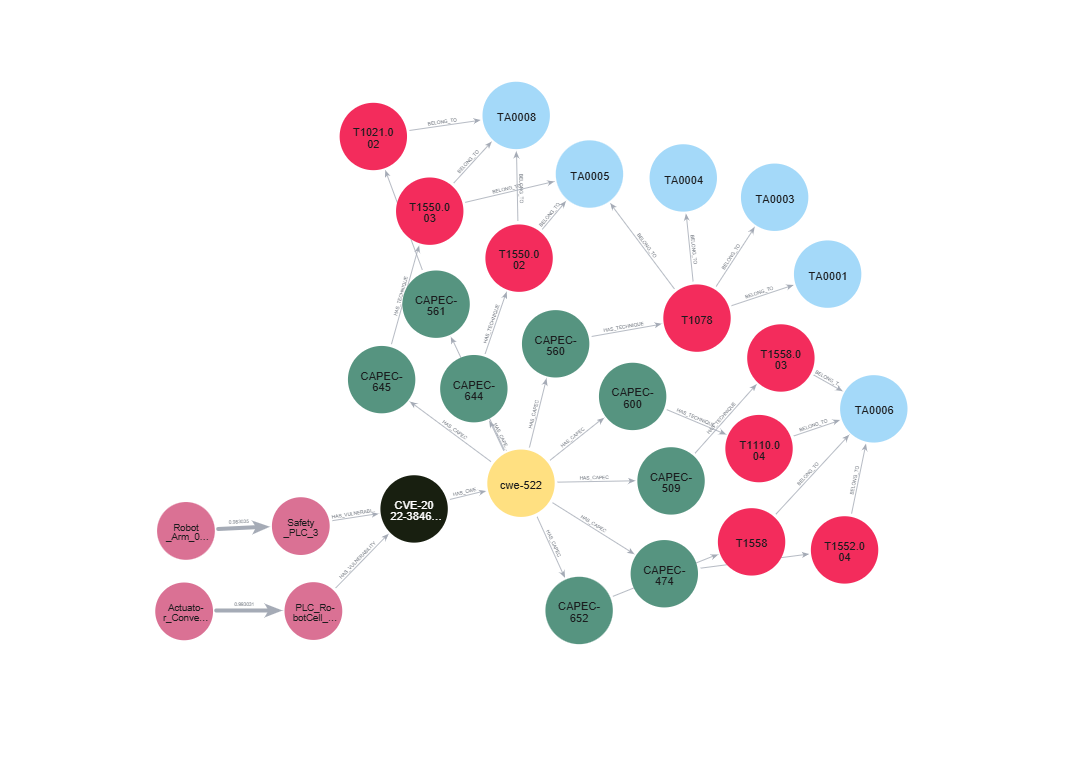}
    \textbf{(a)} High-risk nodes without mitigation links.
\end{minipage}
\hfill
% ---------- Right image (b) ----------
\begin{minipage}{0.48\textwidth}
    \centering
    \includegraphics[width=0.95\linewidth]{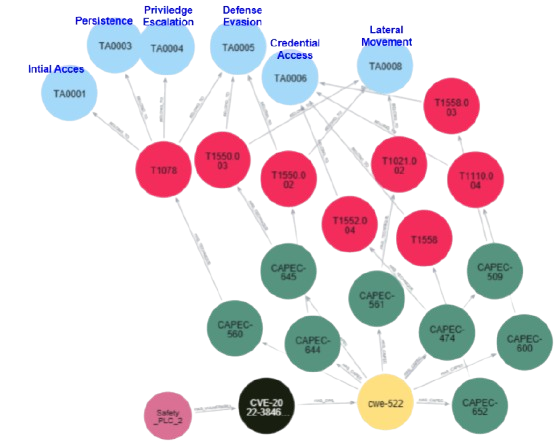}
    \textbf{(b)} Tactic-level attack progression\\
    (Initial Access $\rightarrow$ Lateral Movement).
\end{minipage}
\vspace{4pt}
\caption{Illustrative reasoning use cases of the BRIDG-ICS Knowledge Graph.}
\label{fig:kg-usecases}

\end{figure*}

The potential traversal paths evolve across configurations. In the original dataflow, an 
attacker could progress along 
\(\text{Broker} \rightarrow \text{MES Server} \rightarrow \text{DB} \rightarrow 
\text{Robots} \rightarrow \text{Safety\_PLC}\), leveraging intermediate products to 
reach safety-critical controllers. After enrichment, this route becomes shorter and more 
direct, shifting to 
\(\text{Broker} \rightarrow \text{MES Server} \rightarrow \text{Web Services} \rightarrow 
\text{Safety\_PLC}\), thereby exposing a faster pathway to the Safety\_PLC. As illustrated 
in Figure~\ref{fig:scenario1}, the controlled configuration isolates critical 
nodes and enforces segmentation boundaries, making the pathway from the Broker to the MES 
Server inaccessible and interrupting any progression toward the Safety\_PLC. Comparable 
structural effects appear in other cases, such as 
\(\text{Safety\_PLC} \rightarrow \text{Actuator}\) and 
\(\text{Reverse Proxy} \rightarrow \text{Database}\), demonstrating the consistent benefit 
of control-layer isolation across IT--OT boundaries.

\subsubsection*{Knowledge Graph Use Cases for Threat Reasoning}

Another key use case enabled by BRIDG-ICS is enhanced threat reasoning across enriched 
graph layers. As illustrated in Figures~\ref{fig:kg-usecases}a and \ref{fig:kg-usecases}b, the framework 
can identify high-risk assets (\texttt{riskWeight}~$>$~0.7) that lack mitigation links, offering 
immediate insight into where defensive measures should be prioritised. Moreover, BRIDG-ICS 
captures an adversary’s movement across MITRE ATT\&CK tactic stages by semantically mapping 
device-level indicators---including CVEs, CWEs, and CAPEC patterns---to higher-level attacker 
objectives. These capabilities highlight how knowledge-graph enrichment supports both 
structural attack-path analysis and semantic interpretation of adversarial behavior.

\begin{table}[htbp]
\centering
\caption{Top inter-product connections ranked by average risk, exploitability, and attack cost.}
\label{tab:interproduct_risk}

\setlength{\tabcolsep}{4pt}
\renewcommand{\arraystretch}{1.1}

\begin{tabular}{|l|l|c|c|c|}
\hline
\textbf{Source} & \textbf{Target} & \textbf{Risk} & \textbf{Exploit Prob.} & \textbf{Attack Cost} \\
\hline
Actuator   & PLC          & 0.993 & 0.484 & 0.81 \\ 
\hline
Broker     & Workstation  & 0.886 & 0.974 & 0.44 \\ 
\hline
Robot      & Service      & 0.878 & 0.301 & 0.36 \\ 
\hline
Sensor     & Actuator     & 0.852 & 0.944 & 0.56 \\ 
\hline
Actuator   & Proxy        & 0.851 & 0.279 & 0.56 \\ 
\hline
\end{tabular}
\end{table}

\subsection{Inter-Product Risk and Exploitability Analysis}

A further application of BRIDG-ICS is the assessment of how risk distributes across 
inter-product communication paths. As shown in Table~\ref{tab:interproduct_risk}, the 
framework identifies the most critical product-to-product relationships based on combined 
risk and exploitability measures. The \textit{Actuator--PLC} pair ranks highest in cumulative 
cost, reflecting the central role of actuator devices as gateways toward safety PLCs. 
Similarly, the \textit{Broker--Workstation} and \textit{Sensor--Actuator} links display high 
exploitability, underscoring the exposure of devices that mediate between IT and OT layers. 
These findings offer practical guidance for prioritising mitigation efforts at points of 
amplified propagation potential.

\begin{table}[t]
\centering
\caption{PageRank and Betweenness centrality comparison before and after enrichment (top 5 nodes).}
\label{tab:centrality_nodes}

\small
\setlength{\tabcolsep}{3pt}
\renewcommand{\arraystretch}{1.15}

\begin{tabular}{|l|l|c|c|c|c|c|c|}
\hline
\textbf{Node} & \textbf{Type} & PR Bfr & PR Aft & $\Delta$PR & Betw B. & Betw A. & $\Delta$Betw \\
\hline
PLC.RobotCell\_3 & PLC 
& 2.205 & 2.078 & -0.127 
& $3.17\times10^{-6}$ 
& $7.40\times10^{-4}$ 
& $7.37\times10^{-4}$ \\ \hline

Robot.Arm\_01\_1 & Robot 
& 2.381 & 2.055 & -0.326 
& $3.54\times10^{-6}$ 
& $9.90\times10^{-4}$ 
& $9.86\times10^{-4}$ \\ \hline

IIoT\_Gateway\_1 & Gateway 
& 1.127 & 1.743 & 0.616 
& $3.42\times10^{-5}$ 
& $2.29\times10^{-3}$ 
& $2.25\times10^{-3}$ \\ \hline

Vision\_PC\_2 & Workstation 
& 1.287 & 1.273 & -0.014 
& $3.40\times10^{-5}$ 
& $7.44\times10^{-4}$ 
& $7.10\times10^{-4}$ \\ \hline

PLC.RobotCell\_2 & PLC 
& 0.953 & 1.261 & 0.309 
& $3.42\times10^{-5}$ 
& $9.97\times10^{-4}$ 
& $9.63\times10^{-4}$ \\ \hline
\end{tabular}
\end{table}

\begin{table}[t]
\centering
\caption{Detected communities and their associated risks.}
\label{tab:community}

\small
\renewcommand{\arraystretch}{1.15}
\setlength{\tabcolsep}{3pt}

\begin{tabular}{|p{1.2cm}|c|p{3cm}|c|c|}
\hline
\textbf{Comm. ID} & \textbf{Size} & \textbf{Key Assets} & \textbf{Risk} & \textbf{Cascade} \\
\hline
C83 & 21 & IIoT\_Gateway\_1, Edge\_Historian\_3 & 30.5 & Y \\ \hline
C82 & 18 & PLC\_RobotCell\_3, Robot\_Arm\_01\_1 & 9.6  & Y \\ \hline
C33 & 17 & HMI\_Terminal\_3, Barcode\_Scanner\_2 & 21.3 & Y \\ \hline
C67 & 13 & PLC\_RobotCell\_2, Actuator\_Valve\_A\_2 & 23.8 & N \\ \hline
C46 & 10 & HMI\_Terminal\_1, Sensor\_TankLevel\_1 & 18.8 & N \\ \hline
\end{tabular}

\end{table}

\begin{table}[t]
\centering
\caption{Residual risk comparison across key industrial assets.}
\label{tab:residualrisk}

\begin{tabular}{|l|c|c|c|c|c|c|}
\hline
\textbf{Product Name} & \textbf{Zone} & \textbf{Raw} & \textbf{Enr.} & \textbf{After} & \textbf{$\Delta$} & \textbf{Red. \%} \\
\hline
RAS & DMZ & 49.6 & 59.4 & 9.8 & -49.6 & 16 \\ \hline
VAP11AC & OT & 60.9 & 56.6 & 7.8 & -48.7 & 14 \\ \hline
Control Logix Modules & OT & 0.0 & 7.8 & 7.8 & 0.0 & 100 \\ \hline
Ind. Edge Devices & OT & 9.8 & 15.7 & 7.8 & -7.9 & 50 \\ \hline
Powercenter 1000 & OT & 7.5 & 13.8 & 7.8 & -6.0 & 57 \\ \hline
Eng. Platforms & OT & 7.3 & 13.7 & 7.8 & -5.9 & 57 \\ \hline
Zelio Soft 2 & OT & 11.1 & 16.7 & 7.8 & -8.9 & 47 \\ \hline
Traffic Analyzer & OT & 30.0 & 31.8 & 7.8 & -24.0 & 25 \\ \hline
WallBox & OT & 23.0 & 26.3 & 7.8 & -18.4 & 30 \\ \hline
Modicon M340 & OT & 181.0 & 152.9 & 7.8 & -145.1 & 5 \\ \hline
myPRO & OT & 19.6 & 23.5 & 7.8 & -15.7 & 33 \\ \hline
\end{tabular}

\end{table}

Across all analyses, the enriched graph reveals hidden dependencies and expands the visible 
attack surface, whereas the application of security controls reduces propagation routes and 
limits systemic exposure. This demonstrates BRIDG-ICS’s capacity to integrate quantitative 
risk modelling with semantic reasoning within an interpretable cyber--physical knowledge graph.
\vspace{10 pt}

\noindent
\begin{minipage}{\linewidth}
  \centering
  \includegraphics[width=1.03\linewidth]{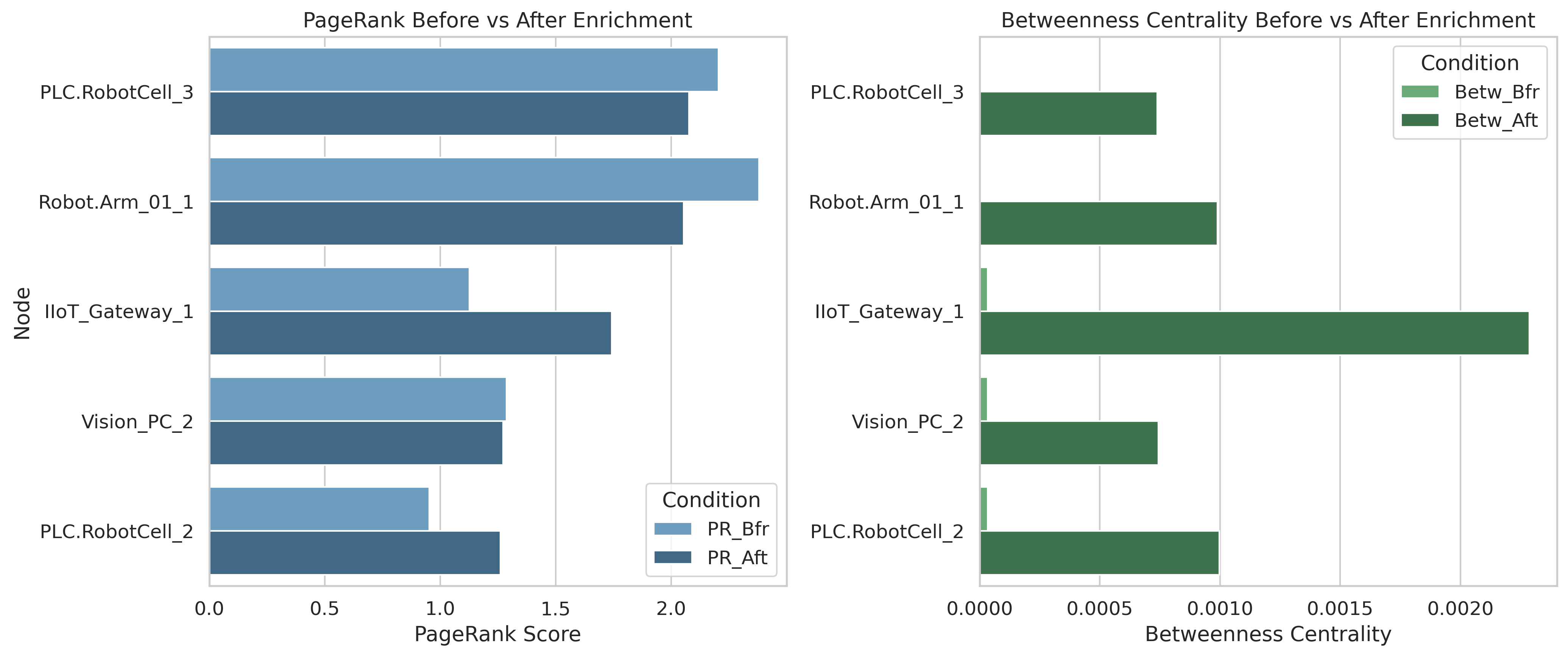}
 \vspace{-3pt}
  \captionof{figure}{Change in node centrality after enrichment (Original $\rightarrow$ Enriched).
  Bars show $\Delta$PageRank and $\Delta$Betweenness for representative assets from Table~\ref{tab:centrality_nodes},
  highlighting redistribution of influence toward communication intermediaries.}
  \label{fig:centrality-delta}
  \vspace{8pt}
\end{minipage}

\section{Discussion}

The results show that BRIDG-ICS successfully integrates heterogeneous cybersecurity and operational datasets into a unified knowledge graph that can model complex cyber–physical threat behaviors. One of the framework’s central contributions is its ability to construct a coherent semantic link between Industry~5.0 architectural knowledge, vulnerability taxonomies (CVE–CWE–CAPEC), and MITRE ATT\&CK adversarial tactics and techniques. This hierarchical integration allows device- and subsystem-level vulnerabilities to be interpreted in context, within the broader landscape of attacker intent and strategy. By bridging operational semantics with adversarial threat models, BRIDG-ICS addresses a key limitation of earlier industrial security knowledge graphs. Prior approaches lacked the cross-domain alignment needed for consistent, context-aware, and intelligence-driven threat analytics.

Another key capability of BRIDG-ICS is its capacity to model \textbf{IT--OT attack propagation}. In the enriched knowledge graph, latent communication dependencies become visible, revealing alternative paths that an attacker might traverse. By contrast, the controlled configuration (with key security controls enforced) shows a measurable reduction in reachable assets and an increase in traversal cost. These observations, reflected in Figures~\ref{fig:overall_propagation} and~\ref{fig:mean_hops_ci}, highlight the value of enrichment for improving attack-surface visibility and demonstrate the efficacy of mitigation controls in constraining lateral movement.

Beyond analyzing individual attack paths, BRIDG-ICS also supports \textbf{system-level structural analysis} to examine how influence and vulnerability are distributed across the industrial environment. It quantifies each node’s importance using two metrics: \textit{PageRank} and \textit{betweenness centrality}. PageRank estimates the relative influence of each component based on the volume and significance of its incoming connections. This highlights devices that can disproportionately shape an attacker’s movement through the network. Betweenness centrality measures how often a node lies on shortest paths between other nodes. Assets with high betweenness act as critical transit points, meaning that compromising them could enable extensive lateral movement.

\begin{figure*}[!ht]
  \vspace{-4pt}
  \centering
  \includegraphics[width=\textwidth]{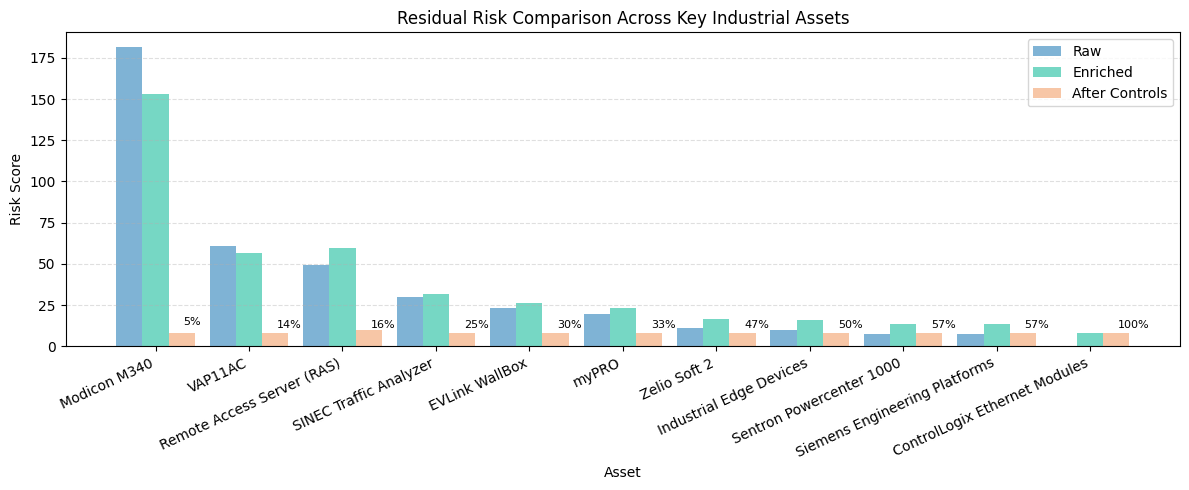}
  \vspace{-6pt}
  \caption{Residual risk comparison across selected industrial assets before enrichment (Raw), after knowledge graph enrichment (Enriched), and after control enforcement (After Controls). Percent labels indicate reported reduction from Table~10.}
  \label{fig:residual-risk}
  \vspace{-4pt}
\end{figure*}

Centrality analysis results (Table~\ref{tab:centrality_nodes} and Figure~\ref{fig:centrality-delta}) indicate that knowledge graph enrichment redistributes structural influence toward communication intermediaries such as gateways, robot-cell PLCs, and workstation nodes\footnote{\href{https://github.com/ahmadspm/Industry-5.0--Intelligent-Threat-Analytics-KGs-and-LLMs/blob/main/Threat-Analysis/page_rank_centrality.txt}{\textit{PageRank Centrality Output (GitHub Repository)}}}. This shift provides a more realistic depiction of operational dependencies and helps identify potential chokepoints. Community detection results (Table~\ref{tab:community}) further highlight tightly coupled subnetworks with elevated cascade potential, emphasising the systemic risks inherent in highly integrated industrial environments\footnote{\href{https://github.com/ahmadspm/Industry-5.0--Intelligent-Threat-Analytics-KGs-and-LLMs/blob/main/Threat-Analysis/community.txt}{\textit{Community Detection Output (GitHub Repository)}}}. Similarly, residual-risk evaluation (Table~\ref{tab:residualrisk} and Figure~\ref{fig:residual-risk}) shows that knowledge graph enrichment increases visibility of interdependencies, while security controls substantially reduce overall exposure. Collectively, these analyses demonstrate that BRIDG-ICS supports not only semantic reasoning and attack-path analysis but also holistic resilience assessment\footnote{\href{https://github.com/ahmadspm/Industry-5.0--Intelligent-Threat-Analytics-KGs-and-LLMs/blob/main/Threat-Analysis/residual_risk.txt}{\textit{Residual Threats Analysis (GitHub Repository)}}}.

Overall, BRIDG-ICS provides a comprehensive foundation for modeling adversarial behavior, identifying vulnerable communication structures, and assessing the impact of defensive controls in smart manufacturing systems. However, some limitations remain. First, it relies on synthetic logs and controlled environments that may not fully capture end-to-end operational variability, though it is still effective for such complex environments. Second, the current enrichment uses only selected knowledge extraction and link-prediction methods. More expressive approaches, such as specialized cyber threat intelligence knowledge graphs or LLM-based predictors, could also be used to uncover additional latent relationships. Third, consolidating extracted artifacts into generic extit{Entity} and extit{DetailAct} classes simplified experimentation, but this abstraction leaves room for the future development of more advanced ontological systems.

The future will extend BRIDG-ICS with \textbf{LLM-assisted reasoning} to enable explainable threat interpretation, adaptive mitigation guidance, and scenario-based simulation. Integrating domain-specialized LLMs with the KGs will support dynamic threat-hypothesis generation, contextual risk prioritization, and automated validation of defensive actions. Building on recent \textbf{Agentic AI} advances, BRIDG-ICS can evolve into a multi-agent analytical ecosystem in which autonomous agents coordinate continuous threat monitoring, attack-path exploration, control-policy evaluation, and real-time what-if analysis. Guided by safety constraints and grounded in the KG, these agentic capabilities can provide proactive defense strategies, strengthen decision support for Industry~5.0 operators, and improve resilience across interconnected industrial systems.

\section{Conclusion}

This work introduced BRIDG-ICS, a knowledge graph framework for modeling and analyzing cyber–physical risks in smart manufacturing. 
By combining contextual enrichment, inferred relations, and probabilistic attack simulation, the framework provides a unified view of how vulnerabilities propagate across IT–OT environments. 
Using domain-specific embeddings such as \textit{CySecBERT} and \textit{SecRoBERTa}, BRIDG-ICS reconstructed missing CVE–CWE–Technique–Tactic links, enhancing semantic completeness and enabling graph-based threat reasoning.

Evaluations across 15 simulated scenarios showed that enrichment reduced average attack path lengths by 20–40\%, while applying NIST-aligned controls increased them by about 25–50\%, demonstrating effective risk mitigation and improved resilience. 
Graph analytics further highlighted critical assets, cascade-prone communities, and measurable post-control risk reductions. BRIDG-ICS unifies knowledge representation and quantitative cyber-risk modeling in a single graph framework. Future work will integrate it with large language models (LLMs) to enable grounded, context-aware threat reasoning and explainable mitigation support for industrial cybersecurity.

\section*{Declarations}

\subsection*{Data Availability}
The datasets used in this study are publicly accessible via the project’s GitHub repository:
\href{https://github.com/ahmadspm/Industry-5.0--Intelligent-Threat-Analytics-KGs-and-LLMs}{\textit{Industry 5.0 Threat Analytics KGs and LLMs Repository}}.
The proprietary local dataset employed for internal testing is not available due to confidentiality constraints.

\subsection*{Funding}
This work was supported by ECU School of Science through Cat 2/3 funding.

\subsection*{Acknowledgements}
We acknowledge the Centre for Securing Digital Futures (CSDF) for providing access to the Industry 5.0 systems testbed and GPU resources used for LLM-driven knowledge graph enrichment.

%% Loading bibliography database
% \bibliography
% \bibliography{refernces}

\begin{appendices}
\renewcommand{\thefigure}{A\arabic{figure}}
\setcounter{figure}{0}
\renewcommand{\thetable}{A\arabic{table}}
\setcounter{table}{0}

\section*{Appendix A. Calculation of Control and Risk Metrics}
\label{ap:calculation}

This appendix provides the definitions of the quantitative parameters
$\{a,c,e,h\}$ and the algorithmic procedure used to compute
\texttt{controlStrength}, $p_{\textit{Exploit}}$,
\texttt{attackCost}, and \texttt{riskWeight} as formalised in
Section~4 (Eqs.~\eqref{eq:controlStrength}--\eqref{eq:riskWeight}).

\subsection*{A.1 Factor Definitions}

The factors are introduced conceptually in Section~4.3.1 and
instantiated here with example computations.

\paragraph{Accessibility ($a$).}
Accessibility quantifies exposure due to authentication strength and
client diversity and is used directly in
\texttt{controlStrength} (Eq.~\eqref{eq:controlStrength}).
For example, in 10{,}000 sessions, 300 were anonymous ($0.03$),
50 used \texttt{security\_mode=None} ($0.005$), and 10 distinct
client IPs were observed, yielding $a \approx 0.03$.

\paragraph{Configuration Hygiene ($c$).}
Configuration hygiene reflects the prevalence of misconfigurations and
the adoption of secure mechanisms. It contributes multiplicatively to
\texttt{controlStrength} in Eq.~\eqref{eq:controlStrength}.
Given a misconfiguration rate of $0.02$, certificate authentication at
$0.90$, and $1\%$ of endpoints failing configuration checks, the
computed value is $c \approx 0.04$.

\paragraph{Exploitability Resistance ($e$).}
Exploitability resistance captures how operational behavior reduces
adversarial opportunity and influences both
\texttt{controlStrength} and the exploit probability
(Eq.~\eqref{eq:pExploit}). With $5\%$ failed writes,
$1\%$ \texttt{AuditWrite} events, and $a \approx 0.03$, one obtains
$e \approx 0.03$.

\paragraph{Hardening Residual Weakness ($h$).}
Residual weakness represents configuration gaps remaining after
hardening and appears in the control-strength term of
Eq.~\eqref{eq:controlStrength}. From certificate usage of $0.90$,
insecure modes at $0.005$, and $1\%$ of endpoints failing security
checks, the resulting value is $h \approx 0.05$.

\subsection*{A.2 Algorithmic Computation Flow}

\begin{samepage}
\small

Algorithm~\ref{alg:risk-computation} computes BRIDG-ICS KG edge-level risk parameters used in KG-driven threat analysis.  
It operationalises Eqs.~\eqref{eq:controlStrength}--\eqref{eq:riskWeight}, enabling quantitative modelling of attack likelihood and propagation across KG communication edges.

\begin{algorithm}[H]
\DontPrintSemicolon
\SetAlgoLined
\caption{Computation of KG Edge-Level Risk Metrics}
\label{alg:risk-computation}

\KwIn{
Operational logs (auth, config, failures);\;
vulnerability data (CVSS, EPSS);\;
asset criticality.
}

\KwOut{
$\texttt{controlStrength}(u,v)$,\;
$p_{\textit{Exploit}}(u,v)$,\;
$\texttt{attackCost}(u,v)$,\;
$\texttt{riskWeight}(u,v)$.
}

\textbf{Step 1: Factor derivation (KG log features):}\;
Compute $a,c,e,h$ from log-derived access strength, configuration hygiene, exploitability resistance, and residual hardening.

\textbf{Step 2: Control Strength (Eq.~\eqref{eq:controlStrength}):}\;
$\texttt{controlStrength}=a\cdot c\cdot e\cdot h.$

\textbf{Step 3: Exploit Probability (Eq.~\eqref{eq:pExploit}):}\;
$p_{\textit{Exploit}}=\texttt{EPSS}\,(1-\texttt{controlStrength}).$

\textbf{Step 4: Attack Cost (Eq.~\eqref{eq:attackCost}):}\;
$\texttt{attackCost}=\text{Base}+f_{AC}+f_{AV}+\texttt{EPSS}.$

\textbf{Step 5: Residual KG Risk (Eq.~\eqref{eq:riskWeight}):}\;
$\texttt{riskWeight}=p_{\textit{Exploit}}\cdot\dfrac{\texttt{criticality}}{10}.$

\Return all KG edge-level threat metrics.

\end{algorithm}

\vspace{-4pt}

Algorithm~\ref{alg:synthetic} generates baseline and secured synthetic logs aligned with the BRIDG-ICS KG topology.  
These logs support KG-based threat modelling by enabling recomputation of Eqs.~\eqref{eq:controlStrength}--\eqref{eq:riskWeight} to assess mitigation effects on risk propagation.

\begin{algorithm}[H]
\DontPrintSemicolon
\SetAlgoLined
\caption{Synthetic Industrial Log Generation for KG Threat Analysis}
\label{alg:synthetic}

\KwIn{
CONTEXT traces;\;
testbed topology and KG-product mappings;\;
security-control profiles.
}

\KwOut{
Baseline and secured OPC-UA + sensor logs for KG risk recomputation.
}

\textbf{Step 1: Baseline extraction:}\;
Derive nominal OPC-UA and sensor patterns from CONTEXT dataset.

\textbf{Step 2: KG topology mapping:}\;
Assign reads/writes and sensor channels to KG \textit{Product} nodes, protocols, and dataflow edges.

\textbf{Step 3: Log synthesis:}\;
Generate baseline logs and secured logs reflecting segmentation, certificates, and patching.

\textbf{Step 4: Threat-metric recomputation:}\;
Recompute $\texttt{controlStrength}$,\,$p_{\textit{Exploit}}$,\,$\texttt{attackCost}$,\,$\texttt{riskWeight}$
via Eqs.~\eqref{eq:controlStrength}--\eqref{eq:riskWeight} to update KG threat scores.

\end{algorithm}

\end{samepage}

\subsection*{C. Attack Scenarios Simulations}
\label{ap:att-scenario}

\begin{figure*}[t]
  \centering
  \captionsetup[subfigure]{labelformat=parens, justification=centering, font=small}

  % ---- Three columns ----
  \subfloat[Original]{%
    \includegraphics[width=0.30\textwidth,trim=30 20 30 20,clip]{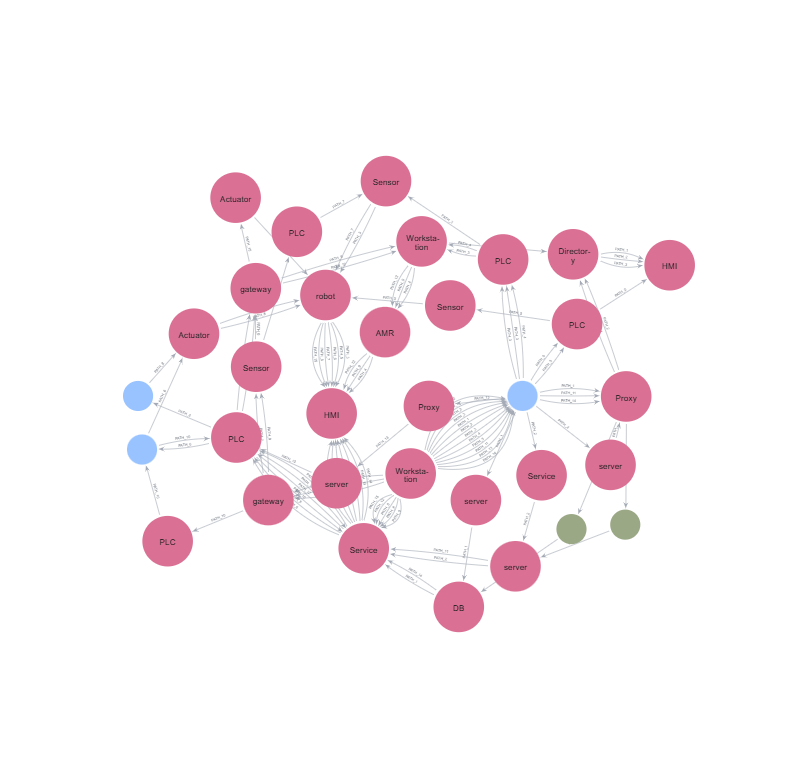}
  }\hfill
  \subfloat[Enriched]{%
    \includegraphics[width=0.30\textwidth,trim=30 20 30 20,clip]{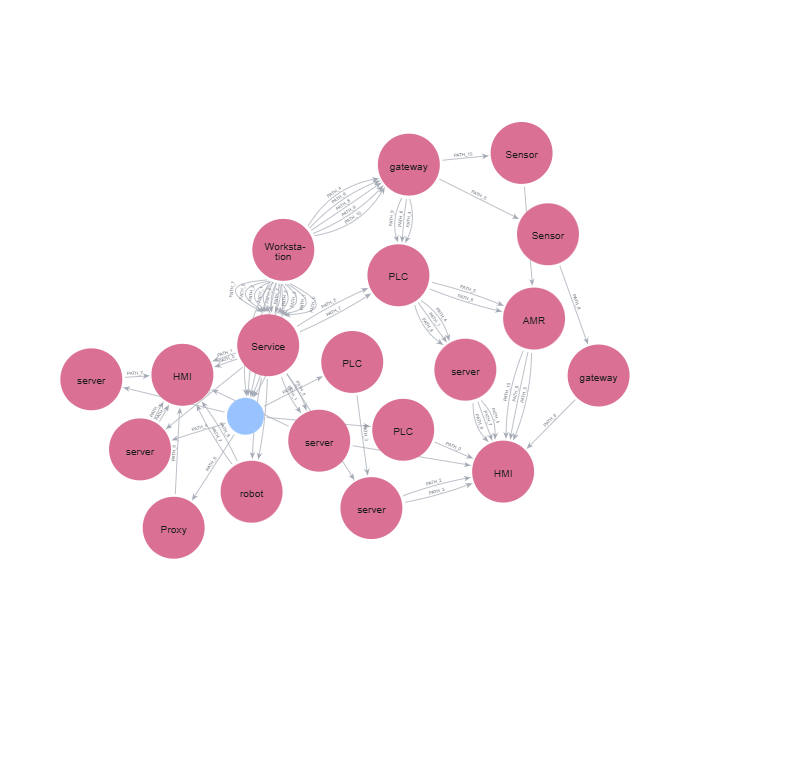}
  }\hfill
  \subfloat[Controlled]{%
    \includegraphics[width=0.30\textwidth,trim=30 20 30 20,clip]{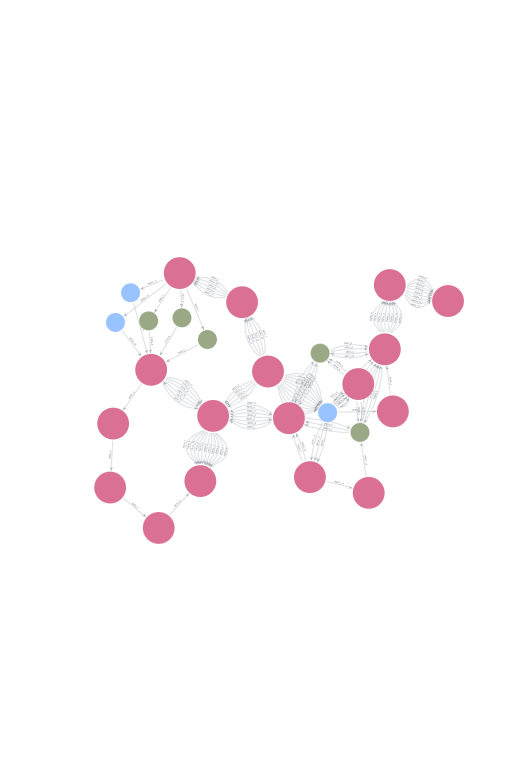}
  }

  \caption{Scenario 2 — Quality Station $\to$ HMI: topology comparison (Original, Enriched, Controlled).}
  \label{fig:scenario2-topology}
\end{figure*}

\begin{figure*}[t]
\centering

% Tight spacing around the figure
\setlength{\abovecaptionskip}{2pt}
\setlength{\belowcaptionskip}{2pt}
\setlength{\floatsep}{4pt}
\setlength{\textfloatsep}{4pt}

% Image trimming (optional—reduces PNG margins)
\newcommand{\sThreeTrim}{trim=20 15 20 15,clip}

\captionsetup[subfigure]{labelformat=parens, justification=centering, font=small}

% ----------- Three columns -----------
\begin{minipage}{0.31\textwidth}
    \centering
    \vspace{24pt}
\includegraphics[
    width=\linewidth,
    trim={20pt 15pt 20pt 15pt},
    clip
]{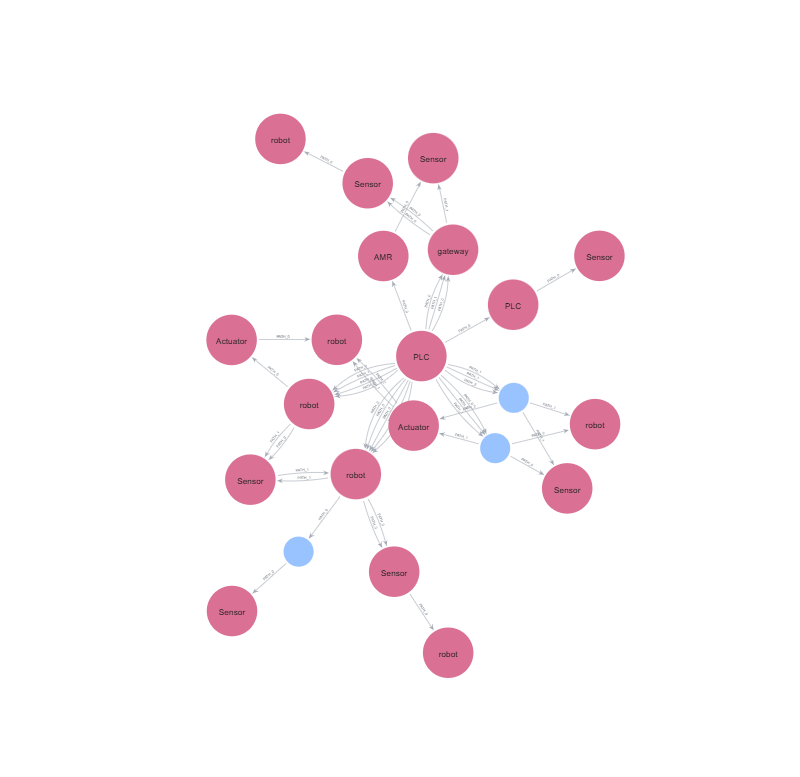}\\[-1pt]

    \vspace{40pt}
    {\small \textbf{(a)} Original}
\end{minipage}
\hfill
\begin{minipage}{0.31\textwidth}
    \centering
\includegraphics[
    width=\linewidth,
    trim={20pt 15pt 20pt 15pt},
    clip
]{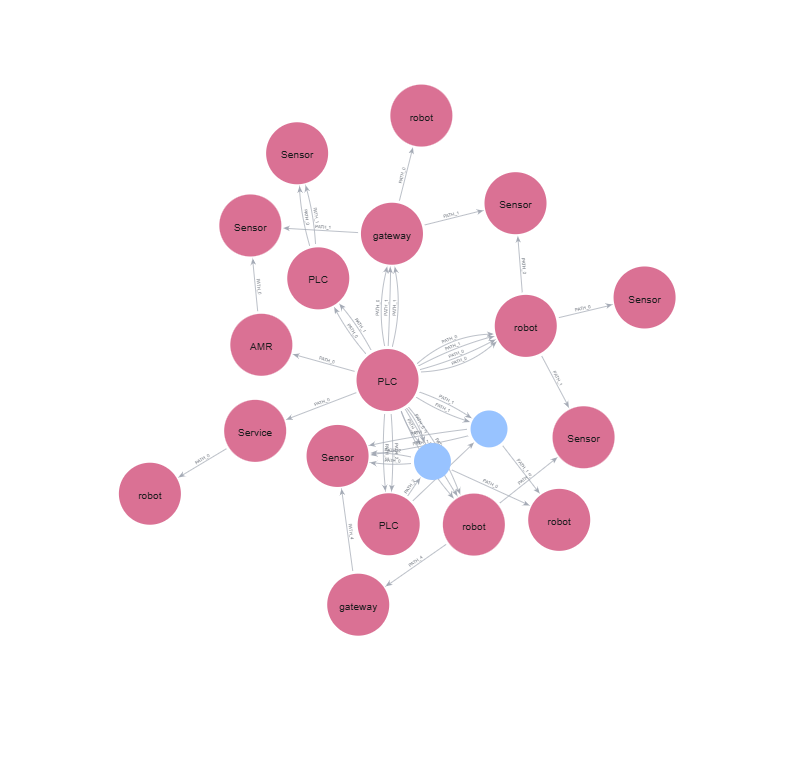}\\[-1pt]

    \vspace{60pt}
    {\small \textbf{(b)} Enriched}
\end{minipage}
\hfill
\begin{minipage}{0.31\textwidth}
    \centering
    \vspace{-3pt}
\includegraphics[
    width=\linewidth,
    trim={20pt 15pt 20pt 15pt},
    clip
]{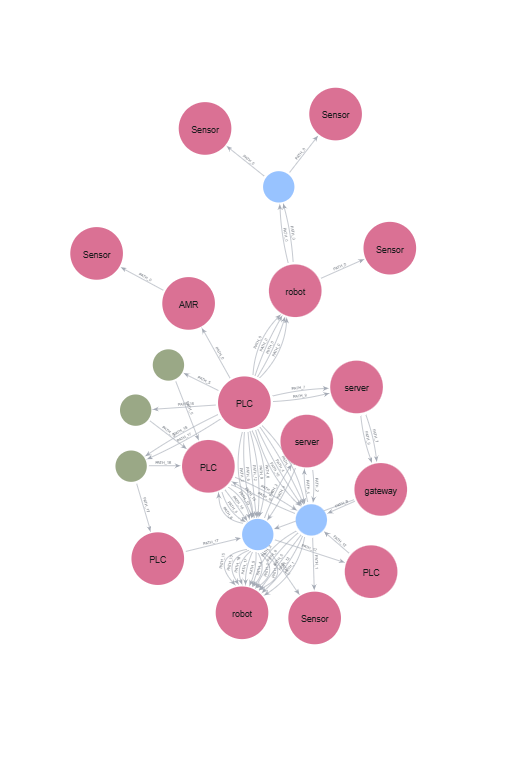}\\[-1pt]

    \vspace{-2pt}
    {\small \textbf{(c)} Controlled}
\end{minipage}

\vspace{2pt}
\caption{Scenario 3 — Safety PLC $\to$ Robot \& Sensor: topology comparison (Original, Enriched, Controlled).}
\label{fig:scenario3}
\end{figure*}

\begin{figure*}[t]
  \centering
  \begin{minipage}{0.30\textwidth}
    \centering
    \includegraphics[height=5cm,keepaspectratio]{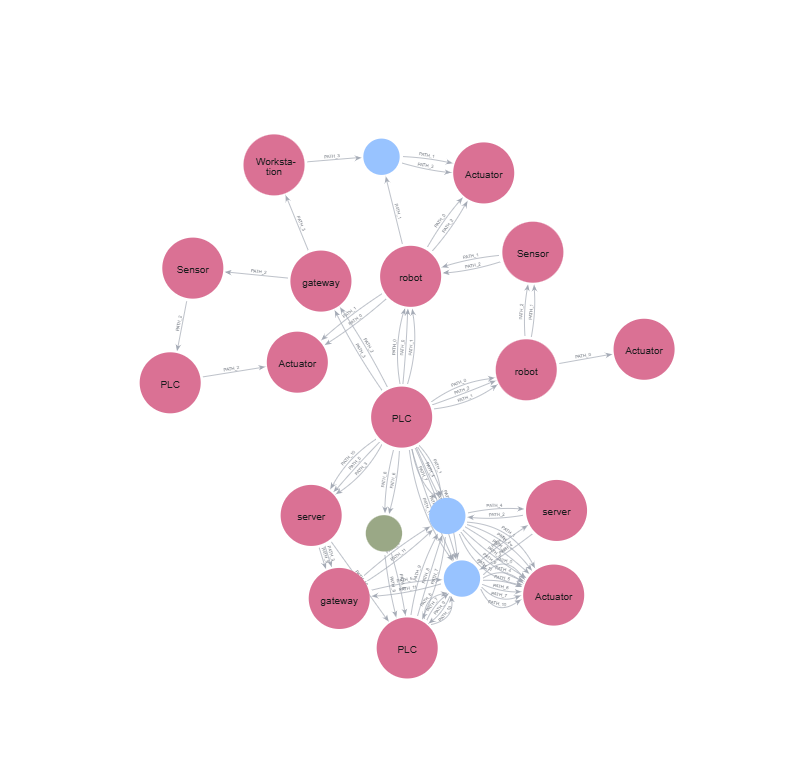}\\
    (a) Original
  \end{minipage}\hspace{0.03\textwidth}%
  \begin{minipage}{0.30\textwidth}
    \centering
    \includegraphics[height=5cm,keepaspectratio]{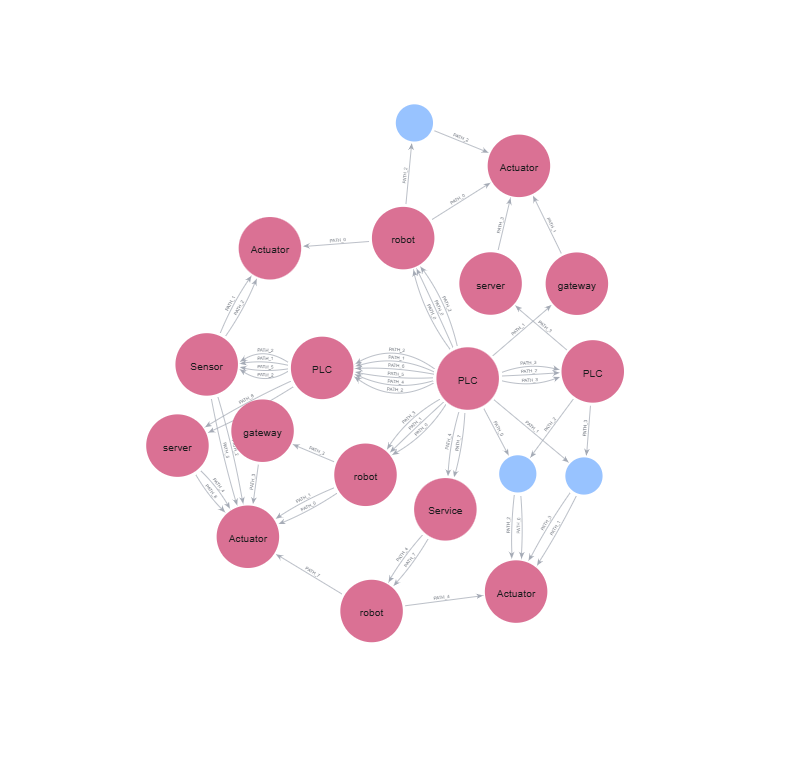}\\
    (b) Enriched
  \end{minipage}\hspace{0.03\textwidth}%
  \begin{minipage}{0.30\textwidth}
    \centering
    \includegraphics[height=4.95cm,keepaspectratio]{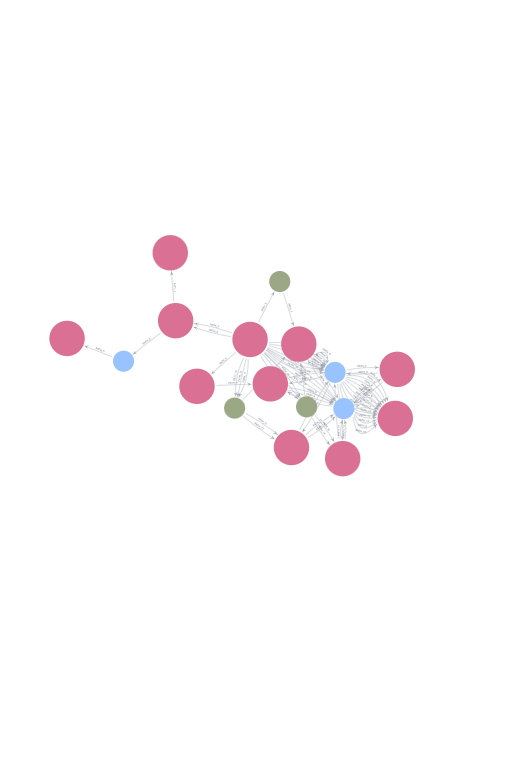}\\
    (c) Controlled
  \end{minipage}

  \caption{Scenario 4 — Safety PLC $\to$ Actuator: topology comparison (Original, Enriched, Controlled).}
  \label{fig:scenario4}
\end{figure*}

\begin{figure*}[t]
  \centering
  \def\panelheight{7cm}

  \begin{minipage}{0.30\textwidth}
    \centering
    \includegraphics[height=\panelheight,keepaspectratio]{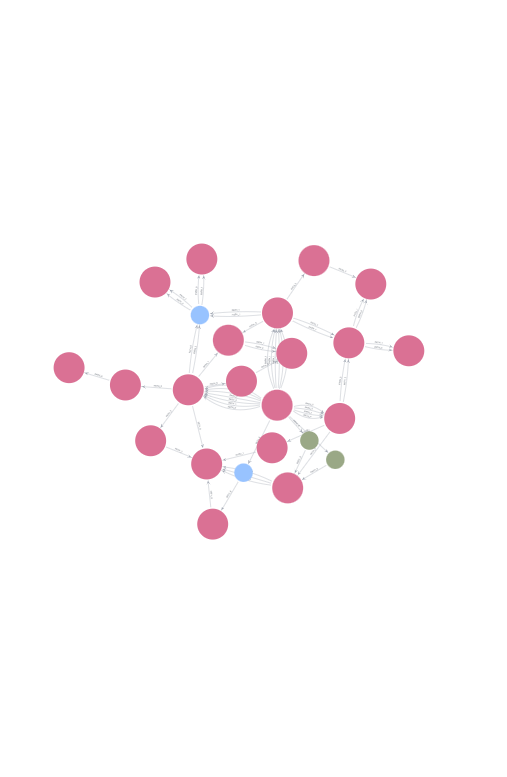}\\
    (a) Original
  \end{minipage}\hspace{0.03\textwidth}%
  \begin{minipage}{0.30\textwidth}
    \centering
    \includegraphics[height=\panelheight,keepaspectratio]{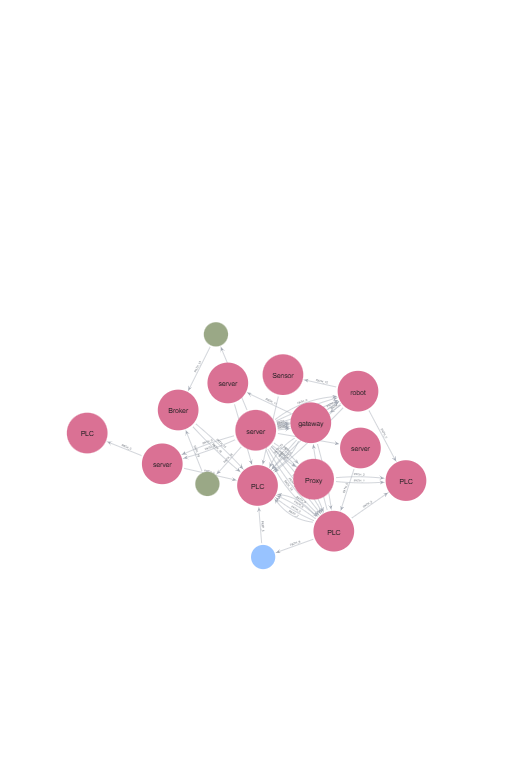}\\
    (b) Enriched
  \end{minipage}\hspace{0.03\textwidth}%
  \begin{minipage}{0.30\textwidth}
    \centering
    \includegraphics[height=\panelheight,keepaspectratio]{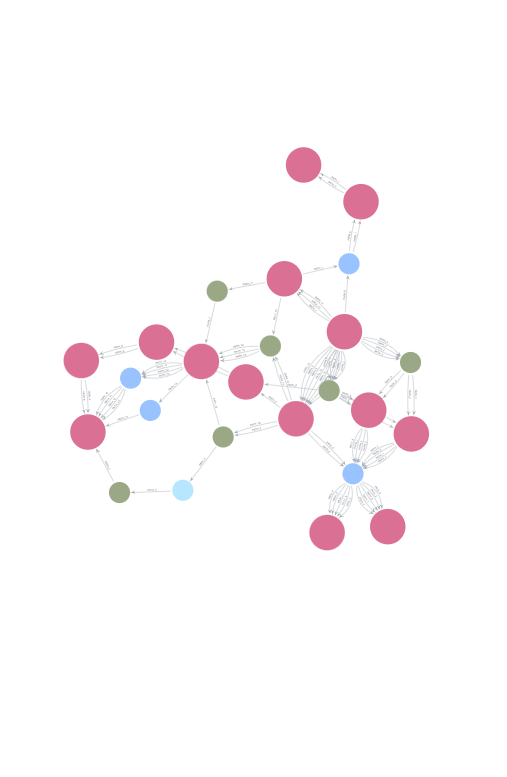}\\
    (c) Controlled
  \end{minipage}
  \vspace{4pt}
  \caption{Scenario 7 — Jump Server $\to$ PLC: topology comparison (Original, Enriched, Controlled).}
  
  \label{fig:scenario7}
\end{figure*}

\begin{figure*}[t]
  \centering
  \begin{minipage}{0.32\textwidth}
    \centering
    \includegraphics[height=5cm,keepaspectratio]{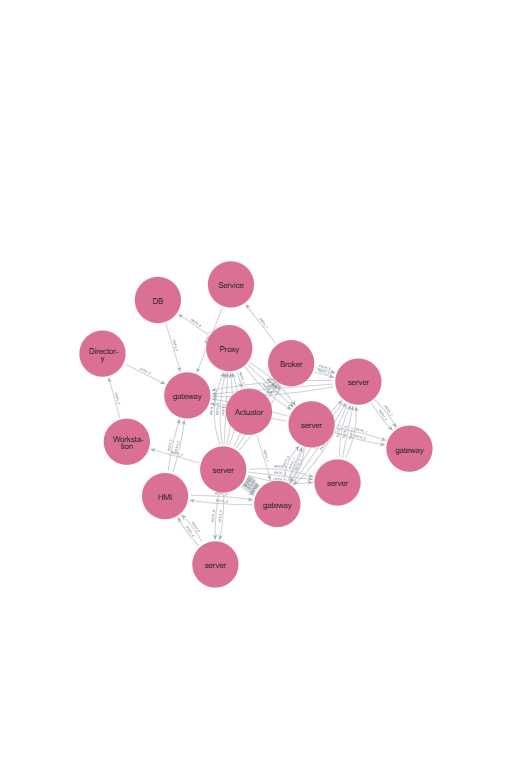}\\
    (a) Original
  \end{minipage}\hfill
  \begin{minipage}{0.32\textwidth}
    \centering
    \includegraphics[height=5cm,keepaspectratio]{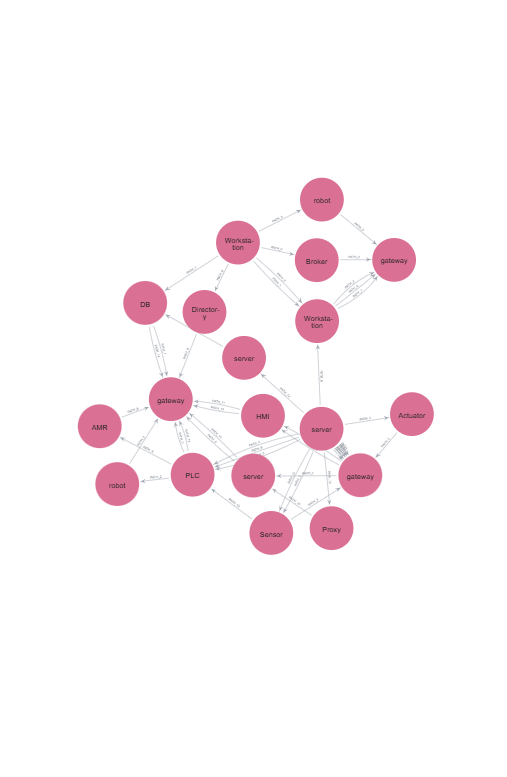}\\
    (b) Enriched
  \end{minipage}\hfill
  \begin{minipage}{0.32\textwidth}
    \centering
    \includegraphics[height=5cm,keepaspectratio]{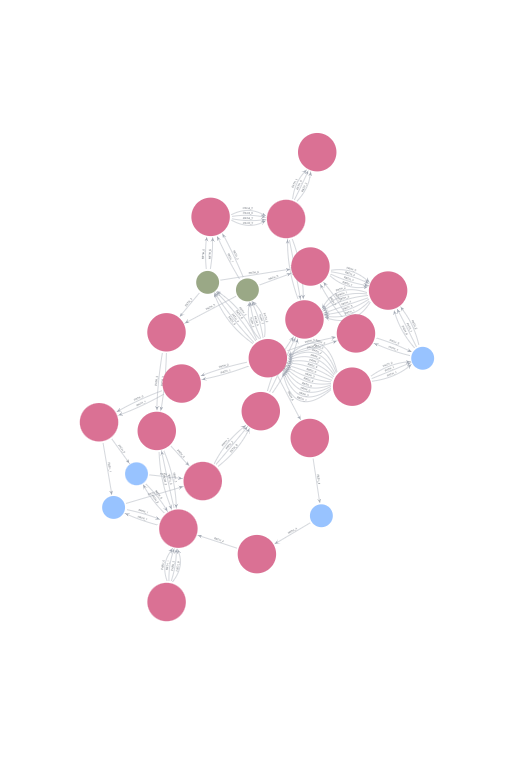}\\
    (c) Controlled
  \end{minipage}
  \vspace{4pt}
  \caption{Scenario 15 — Engineering Workstation / SCADA $\to$ DB: topology comparison (Original, Enriched, Controlled).}
   \label{fig:scenario15-topology}
\end{figure*}

% ---- Text sits on top; table centered below ----
\noindent Table~\ref{tab:industry5-scenarios} presents fifteen simulated cybersecurity attack scenarios used to evaluate the Industry~5.0 Knowledge Graph (KG) threat model. Each scenario highlights adversarial propagation, the effect of enrichment, and resilience under controls within the BRIDG–ICS framework. Figures~\ref{fig:scenario2-topology}–\ref{fig:scenario15-topology} visualize these cases in knowledge engine under the \textit{Original}, \textit{Enriched}, and \textit{Controlled} KG configurations. Figure~\ref{fig:overall_propagation}(a) compares average hop counts, while Figure~\ref{fig:overall_propagation}(b) shows the number of reachable products.

\subsection*{D. Attack Path Analysis Using knowledge engine Graph Data Science}

Attack path analysis was conducted using Neo4j Graph Data Science (GDS) to evaluate potential propagation routes between industrial products. 
In this study, Yen’s \textit{k}-shortest path algorithm (\(k = 20\)) was used to identify multiple plausible attack paths between entry and target products. 
Although Yen’s algorithm operates on unweighted graphs, quantitative attributes such as \texttt{riskWeight}, \texttt{attackScenario}, and \texttt{pExploit} were retained to support subsequent risk interpretation and relation inference—particularly in constructing \texttt{HAS\_POSSIBLE\_COMMUNICATION} links and evaluating propagation likelihoods.

For comparative purposes, a weighted version of the analysis using Dijkstra’s algorithm was also developed, enabling the inclusion of edge weights based on \texttt{riskWeight} for cost-aware path estimation. 
Full Cypher query implementations for both Yen’s and Dijkstra’s algorithms are available in the project repository.\footnote{\url{https://github.com/ahmadspm/Industry-5.0--Intelligent-Threat-Analytics-KGs-and-LLMs/blob/main/Threat-Analysis/Example.txt}}

\begin{table}[t]
\centering
\small
\setlength{\tabcolsep}{4pt}
\renewcommand{\arraystretch}{1.15}

\caption{Summary of Industry~5.0 Cybersecurity Scenarios Evaluated Using the BRIDG-ICS Knowledge Graph}
\label{tab:industry5-scenarios}

\begin{tabular}{|p{0.45\textwidth}|p{0.45\textwidth}|}
\hline
\textbf{Scenario} & \textbf{Key Threat Impact} \\
\hline

S2: Quality Station → HMI.  
Enrichment shortens hops; controls increase distance. &
Credential misuse enables lateral movement and operator manipulation. \\
\hline

S3: Safety PLC → Robot/Sensor.  
Enrichment reduces hop length. &
Logic tampering degrades safety interlocks and sensing. \\
\hline

S4: Safety PLC → Actuator.  
Enrichment expands reach. &
Overflow faults cause unsafe or erratic actuator behavior. \\
\hline

S5: Reverse Proxy → DB.  
Enrichment shortens path. &
SQL injection enables MES data corruption/exfiltration. \\
\hline

S6: Email Gateway → DB.  
Enrichment reduces hops. &
Phishing pivot leaks credentials and enables spread. \\
\hline

S7: Jump Server → PLC.  
Enrichment slightly shortens path. &
Compromise bypasses segmentation and allows PLC command abuse. \\
\hline

S8: Rogue Sensor Injection.  
Enrichment reveals propagation. &
Injected data poisons feedback loops and situational awareness. \\
\hline

S9: Sensor → Robot/Actuator.  
Enrichment increases reach. &
Manipulated sensor signals trigger unsafe robot/actuator actions. \\
\hline

S10: Update Mirrors → PLC.  
Enrichment increases reachability. &
Compromised mirrors push malicious firmware widely. \\
\hline

S11: Jump Server → Service.  
Enrichment finds new routes. &
Unrestricted admin paths expose management APIs. \\
\hline

S12: SCADA → PLC.  
Enrichment reveals new lateral paths. &
SCADA takeover risks system-wide disruption. \\
\hline

S13: Actuator → Robot.  
Enrichment reduces hops. &
Tampering causes misalignment and quality defects. \\
\hline

S14: Sensor → Database.  
Enrichment reduces hop count. &
False QA data affects traceability and quality. \\
\hline

S15: Eng. Workstation/SCADA → Database.  
Enrichment shortens path. &
Privilege escalation leads to MES exfiltration; controls restore segmentation. \\
\hline

\end{tabular}

\end{table}

\vspace{-20pt}
% ---------- Scenario 3 ----------

\vspace{-20pt}
% ---------- Scenario 4 ----------
\vspace{-20pt}

% ---------- Scenario 5 ----------
\begin{figure*}[t]
  \centering
  \begin{minipage}{0.32\textwidth}
    \centering
    \includegraphics[height=5cm,keepaspectratio]{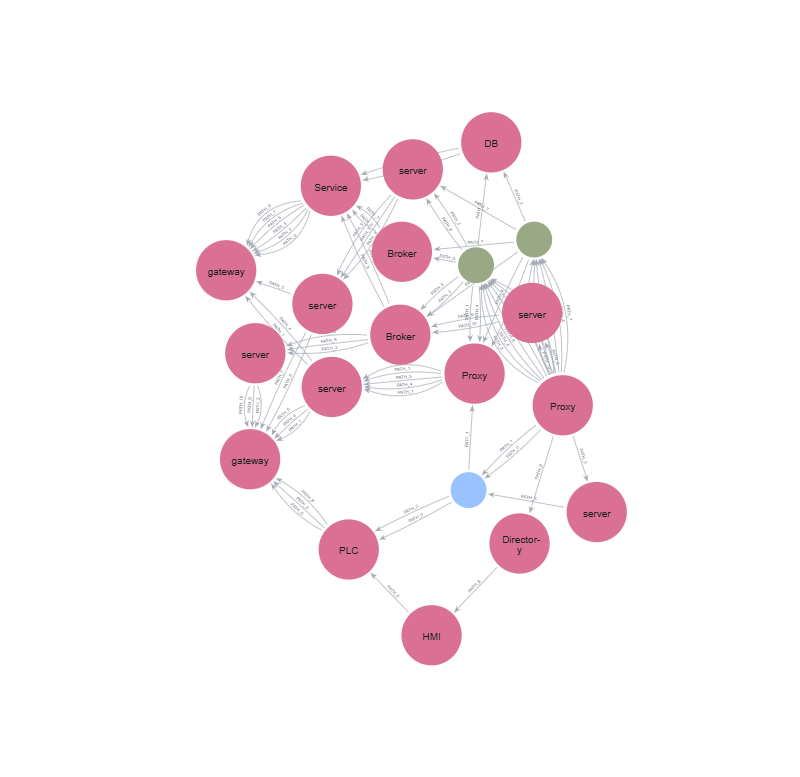}\\
    (a) Original
  \end{minipage}\hfill
  \begin{minipage}{0.32\textwidth}
    \centering
    \includegraphics[height=5cm,keepaspectratio]{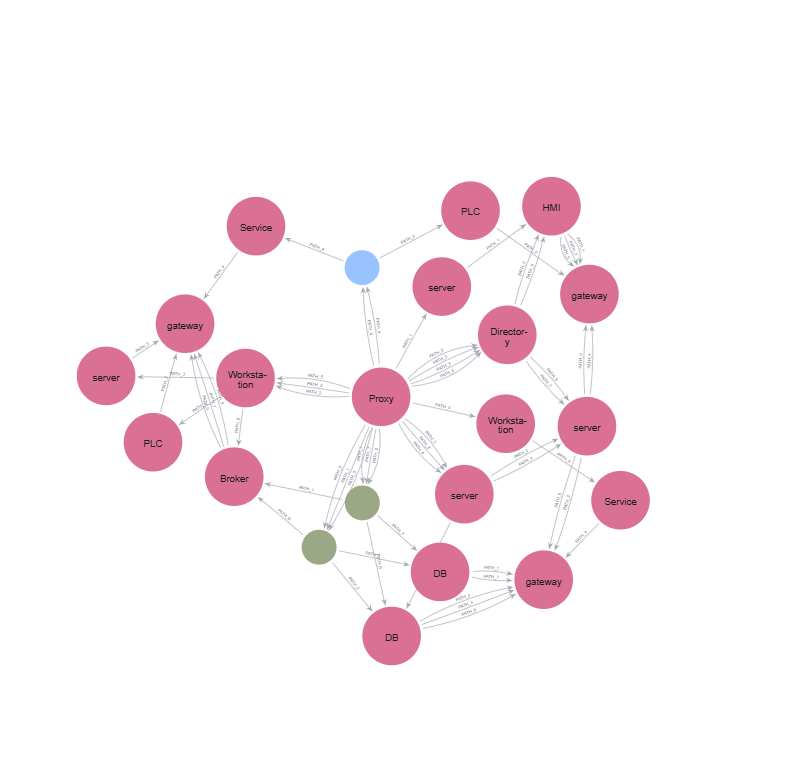}\\
    (b) Enriched
  \end{minipage}\hfill
  \begin{minipage}{0.32\textwidth}
    \centering
    \includegraphics[height=5cm,keepaspectratio]{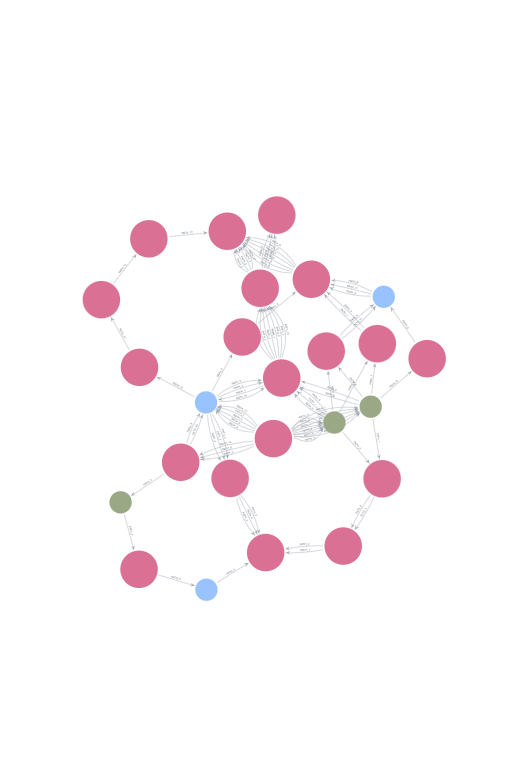}\\
    (c) Controlled
  \end{minipage}
  \caption{Scenario 5 — Reverse Proxy $\to$ DB: topology comparison (Original, Enriched, Controlled).}
\end{figure*}
\vspace{-20pt}
% ---------- Scenario 6 ----------
\begin{figure*}[t]
  \centering
  \begin{minipage}{0.32\textwidth}
    \centering
    \includegraphics[height=5cm,keepaspectratio]{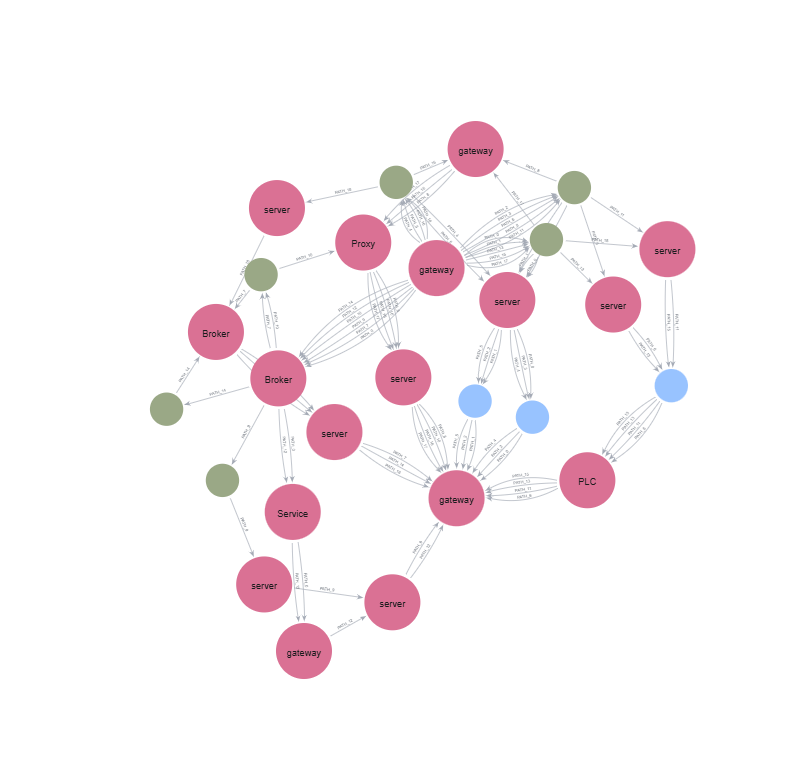}\\
    (a) Original
  \end{minipage}\hfill
  \begin{minipage}{0.32\textwidth}
    \centering
    \includegraphics[height=5cm,keepaspectratio]{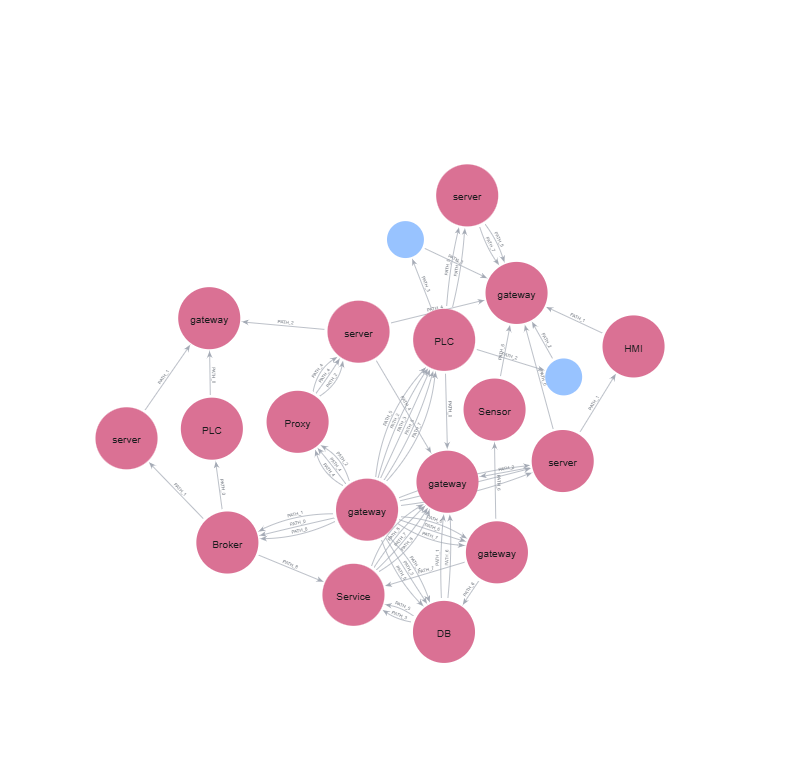}\\
    (b) Enriched
  \end{minipage}\hfill
  \begin{minipage}{0.32\textwidth}
    \centering
    \includegraphics[height=5cm,keepaspectratio]{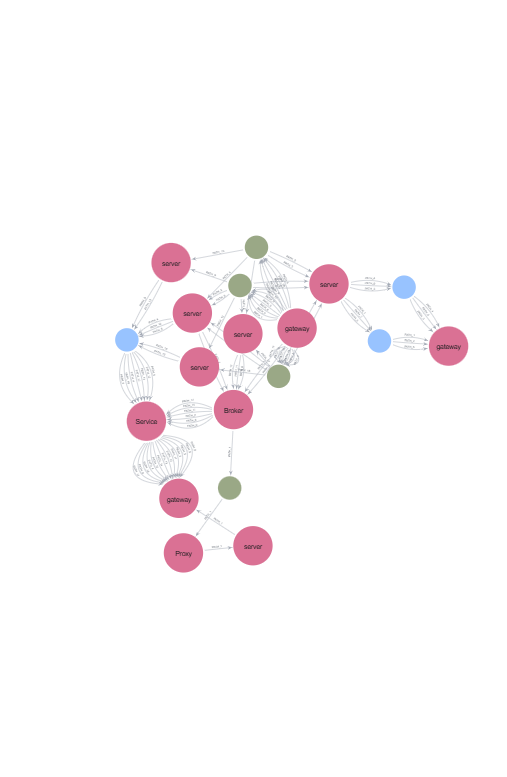}\\
    (c) Controlled
  \end{minipage}
  \vspace{4pt}
  \caption{Scenario 6 — Email Gateway $\to$ DB: topology comparison (Original, Enriched, Controlled).}
\end{figure*}
\vspace{-50pt}
% ---------- Scenario 7 ----------

\vspace{-20pt}
% ---------- Scenario 8 (no controlled) ----------
\begin{figure*}[t]
  \centering
  \begin{minipage}{0.48\textwidth}
    \centering
    \includegraphics[height=5cm,keepaspectratio]{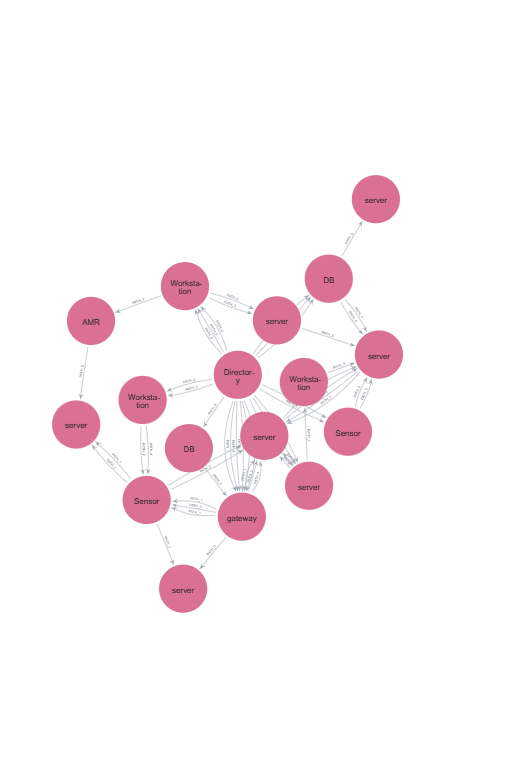}\\
    (a) Original
  \end{minipage}\hfill
  \begin{minipage}{0.48\textwidth}
    \centering
    \includegraphics[height=5cm,keepaspectratio]{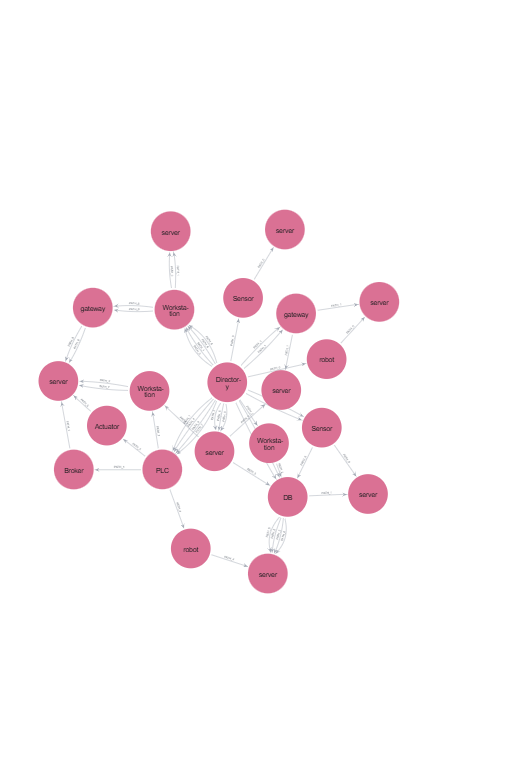}\\
    (b) Enriched
  \end{minipage}
  \caption{Scenario 8 — Rogue Sensor Injection: topology comparison (Original, Enriched).}
\end{figure*}
\vspace{-20pt}
% ---------- Scenario 9 (no controlled) ----------
\begin{figure*}[t]
  \centering
  \begin{minipage}{0.48\textwidth}
    \centering
    \includegraphics[height=5cm,keepaspectratio]{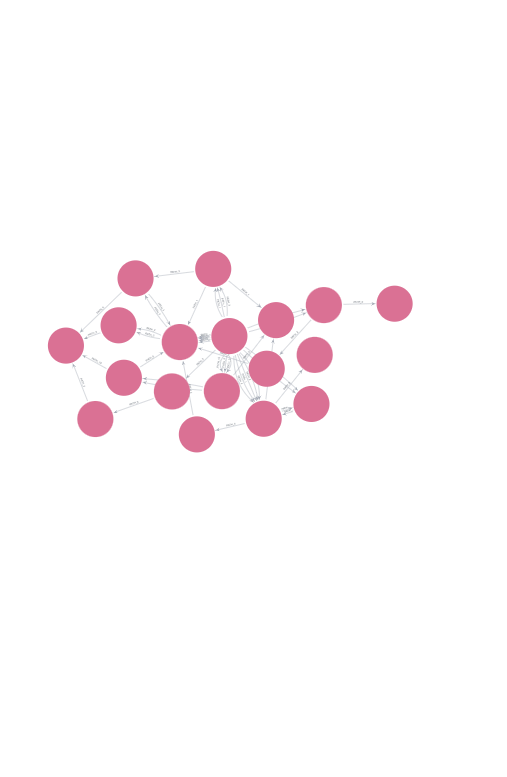}\\
    (a) Original
  \end{minipage}\hfill
  \begin{minipage}{0.48\textwidth}
    \centering
    \includegraphics[height=5cm,keepaspectratio]{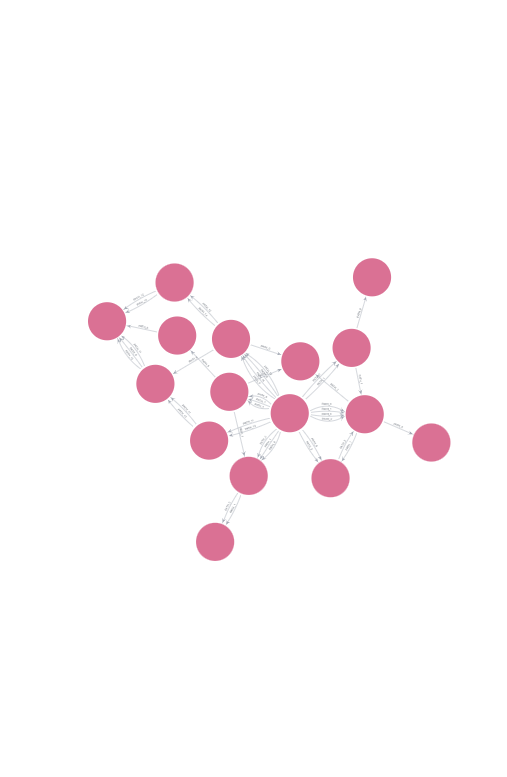}\\
    (b) Enriched
  \end{minipage}
  \vspace{4pt}
  \caption{Scenario 9 — Sensor $\to$ Robot \& Actuator: topology comparison (Original, Enriched).}
\end{figure*}
\vspace{-20pt}
% ---------- Scenario 10 ----------
\begin{figure*}[t]
  \centering
  \begin{minipage}{0.32\textwidth}
    \centering
    \includegraphics[height=5cm,keepaspectratio]{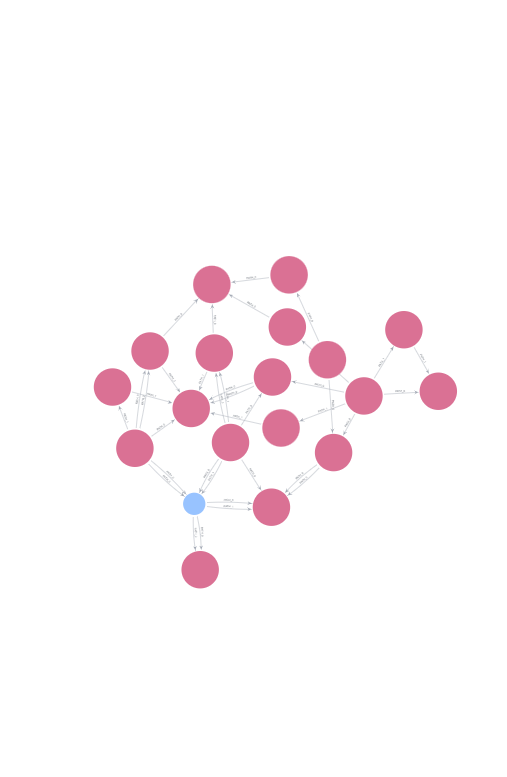}\\
    (a) Original
  \end{minipage}\hfill
  \begin{minipage}{0.32\textwidth}
    \centering
    \includegraphics[height=5cm,keepaspectratio]{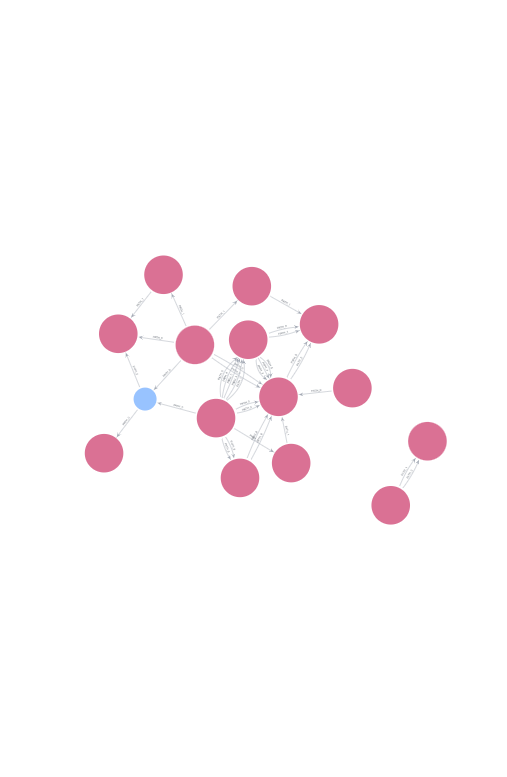}\\
    (b) Enriched
  \end{minipage}\hfill
  \begin{minipage}{0.32\textwidth}
    \centering
    \includegraphics[height=5cm,keepaspectratio]{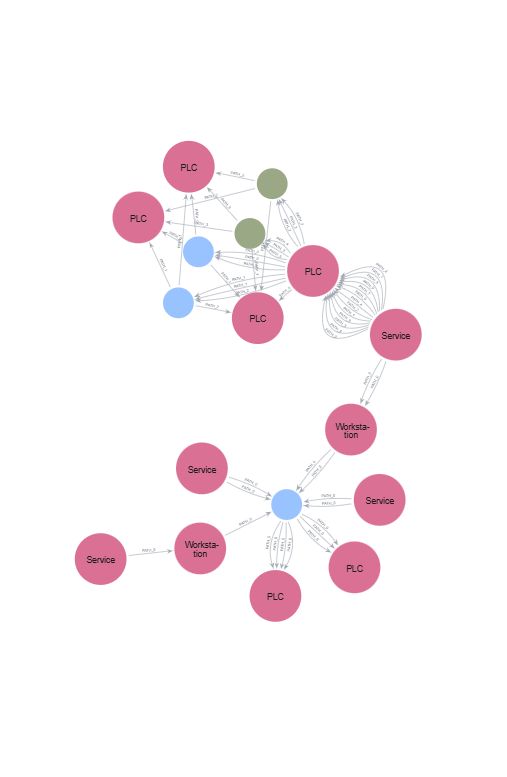}\\
    (c) Controlled
  \end{minipage}
  \vspace{4pt}
  \caption{Scenario 10 — Backup / Update Mirrors $\to$ PLC: topology comparison (Original, Enriched, Controlled).}
\end{figure*}
\vspace{-20pt}
% ---------- Scenario 11 ----------
\begin{figure*}[t]
  \centering
  \begin{minipage}{0.32\textwidth}
    \centering
    \includegraphics[height=5cm,keepaspectratio]{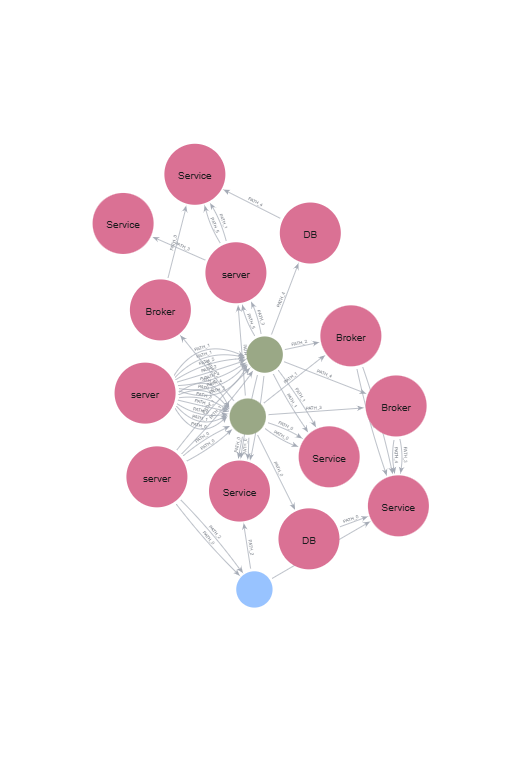}\\
    (a) Original
  \end{minipage}\hfill
  \begin{minipage}{0.32\textwidth}
    \centering
    \includegraphics[height=5cm,keepaspectratio]{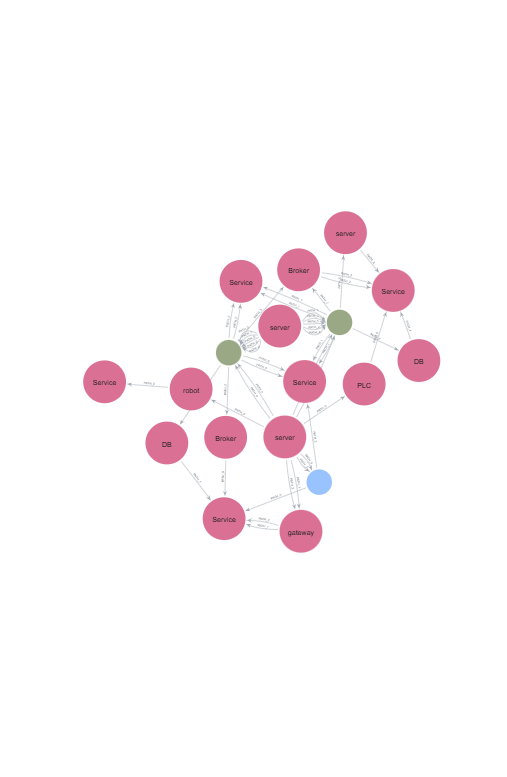}\\
    (b) Enriched
  \end{minipage}\hfill
  \begin{minipage}{0.32\textwidth}
    \centering
    \includegraphics[height=5cm,keepaspectratio]{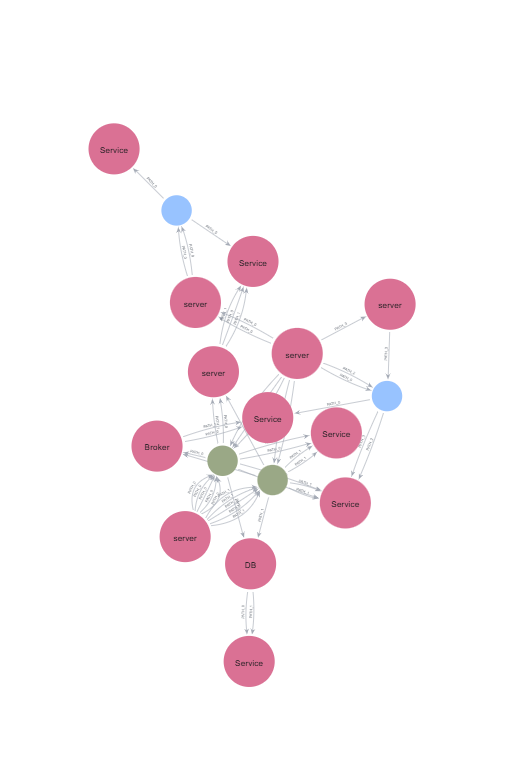}\\
    (c) Controlled
  \end{minipage}
  \vspace{4pt}
  \caption{Scenario 11 — Jump Server $\to$ Service: topology comparison (Original, Enriched, Controlled).}
\end{figure*}
\vspace{-20pt}
% ---------- Scenario 12 ----------
\begin{figure*}[t]
  \centering
  \begin{minipage}{0.32\textwidth}
    \centering
    \includegraphics[height=6.5cm,keepaspectratio,trim=40 40 40 40,clip]{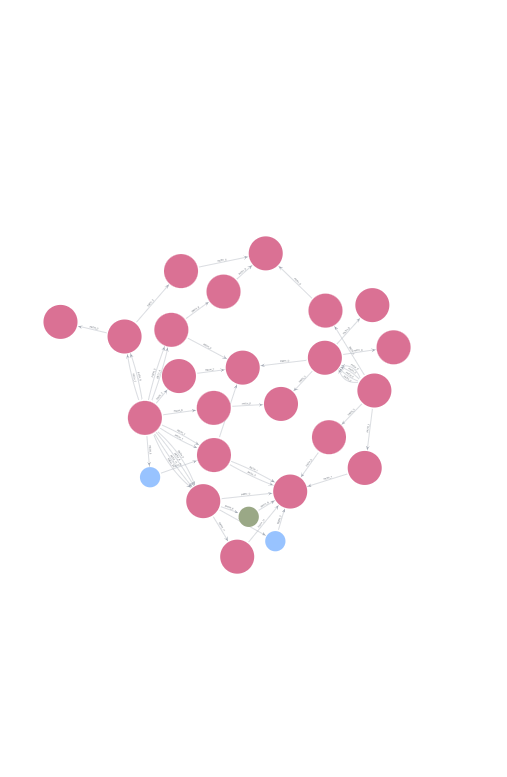}\\[-3pt]
    (a) Original
  \end{minipage}\hfill
  \begin{minipage}{0.32\textwidth}
    \centering
    \includegraphics[height=6.5cm,keepaspectratio,trim=40 40 40 40,clip]{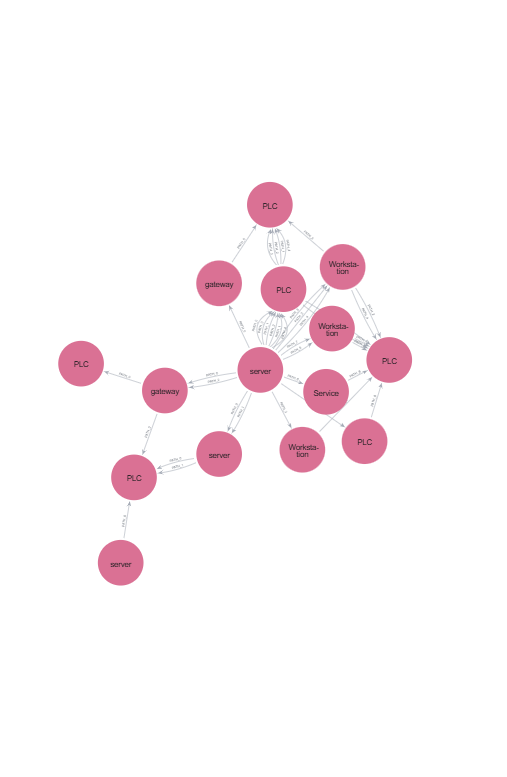}\\[-3pt]
    (b) Enriched
  \end{minipage}\hfill
  \begin{minipage}{0.32\textwidth}
    \centering
    \includegraphics[height=6.5cm,keepaspectratio,trim=40 40 40 40,clip]{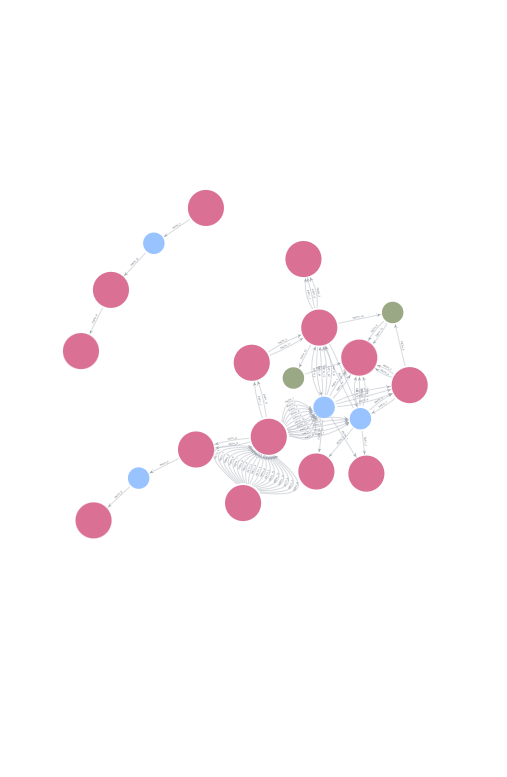}\\[-3pt]
    (c) Controlled
  \end{minipage}
  \vspace{6pt}
  \caption{Scenario 12 — SCADA Server $\to$ PLC: topology comparison (Original, Enriched, Controlled).}
\end{figure*}
\vspace{-20pt}
% ---------- Scenario 13 ----------
\begin{figure*}[t]
  \centering
  \begin{minipage}{0.32\textwidth}
    \centering
    \includegraphics[height=6.5cm,keepaspectratio,trim=40 40 40 40,clip]{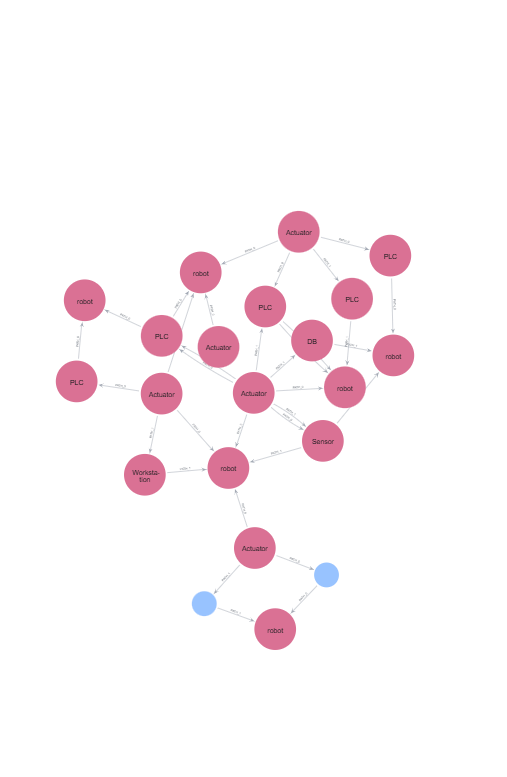}\\[-3pt]
    (a) Original
  \end{minipage}\hfill
  \begin{minipage}{0.32\textwidth}
    \centering
    \includegraphics[height=6.5cm,keepaspectratio,trim=40 40 40 40,clip]{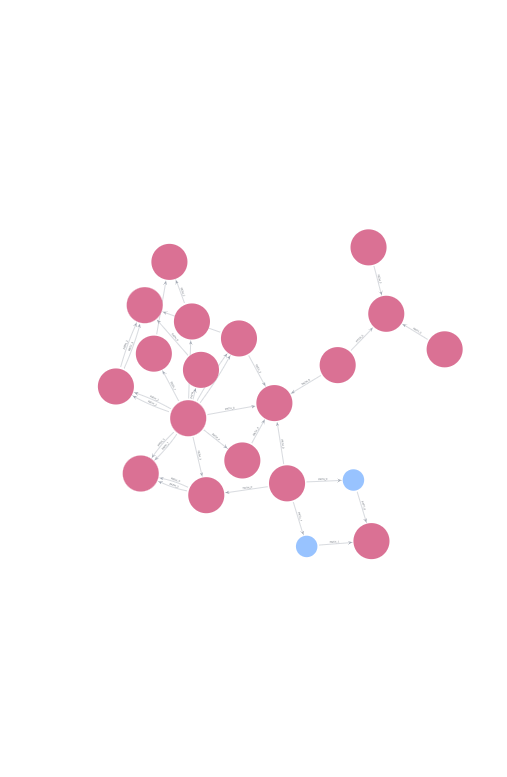}\\[-3pt]
    (b) Enriched
  \end{minipage}\hfill
  \begin{minipage}{0.32\textwidth}
    \centering
    \includegraphics[height=6.5cm,keepaspectratio,trim=40 40 40 40,clip]{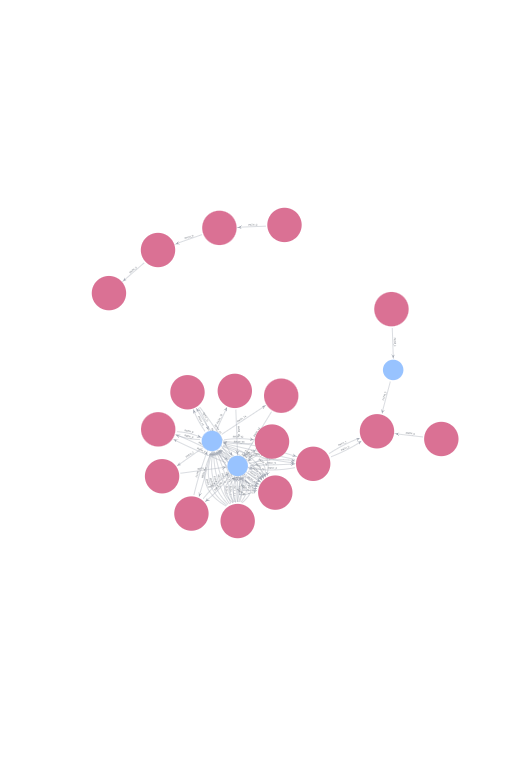}\\[-3pt]
    (c) Controlled
  \end{minipage}
  \vspace{4pt}
  \caption{Scenario 13 — Actuator $\to$ Robot: topology comparison (Original, Enriched, Controlled).}
\end{figure*}
\vspace{-20pt}
% ---------- Scenario 14 ----------
\begin{figure*}[t]
  \centering
  \begin{minipage}{0.32\textwidth}
    \centering
    \includegraphics[height=6.5cm,keepaspectratio,trim=40 40 40 40,clip]{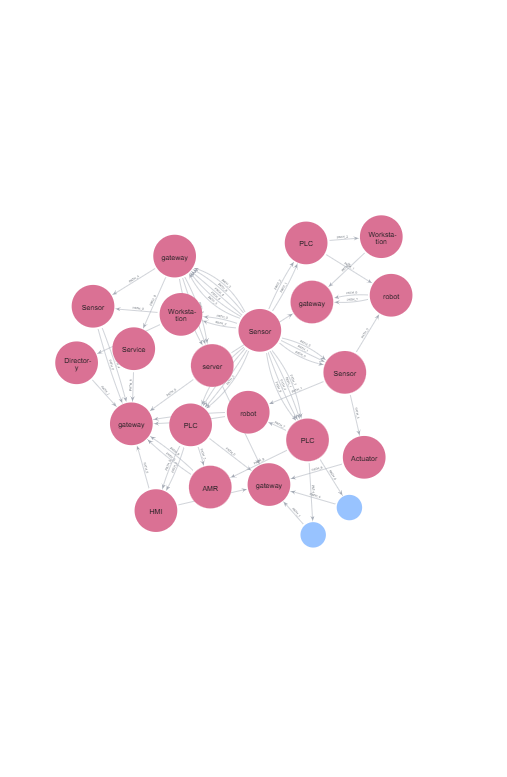}\\[-3pt]
    (a) Original
  \end{minipage}\hfill
  \begin{minipage}{0.32\textwidth}
    \centering
    \includegraphics[height=6.5cm,keepaspectratio,trim=40 40 40 40,clip]{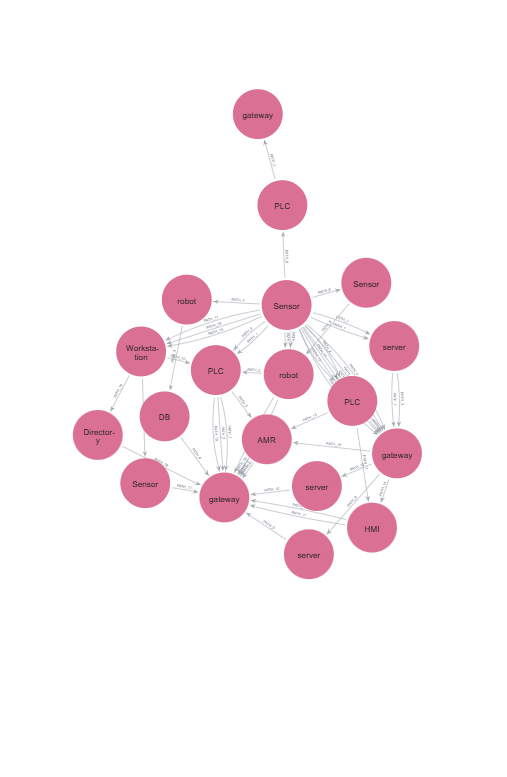}\\[-3pt]
    (b) Enriched
  \end{minipage}\hfill
  \begin{minipage}{0.32\textwidth}
    \centering
    \includegraphics[height=6.5cm,keepaspectratio,trim=40 40 40 40,clip]{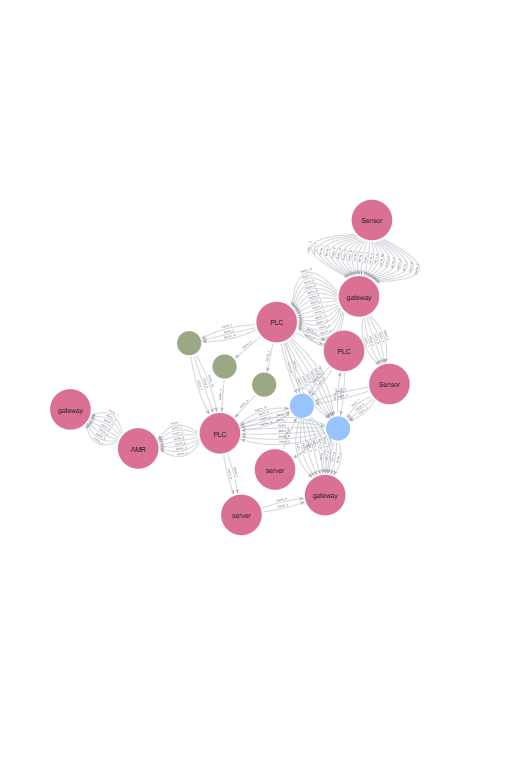}\\[-3pt]
    (c) Controlled
  \end{minipage}
  \vspace{4pt}
  \caption{Scenario 14 — Sensor $\to$ DB: topology comparison (Original, Enriched, Controlled).}
\end{figure*}
\vspace{-20pt}
% ---------- Scenario 15 ----------

\end{appendices}


\end{document}